\documentclass{lmcs} %%% last changed 2014-08-20
\pdfoutput=1

\usepackage{lastpage}
\lmcsdoi{20}{4}{23}
\lmcsheading{}{\pageref{LastPage}}{}{}%
{Feb.~06,~2024}{Dec.~11,~2024}{}


\usepackage{tikz}
\usetikzlibrary{automata}
\usepackage[utf8]{inputenc}
\usepackage{hyperref}
\usepackage{amsmath,amssymb}
\usepackage{graphicx}

\renewcommand{\vec}[1]{\mathbf{#1}}

\newcommand{\setN}{\mathbb{N}}

\newcommand{\prest}{\textsc{Pre}^*}
\newcommand{\postst}{\textsc{Post}^*}
\newcommand{\reach}{\xrightarrow{*}}
\newcommand{\reachA}{\xrightarrow{A^*}}

\newcommand{\post}{\textsc{post}^*_A}
\newcommand{\pre}{\textsc{pre}^*_A}

\newcommand{\dreg}[1]{\downarrow\hspace{-0.2em}#1}
\newcommand{\reg}[1]{\downarrow\hspace{-0.2em}#1}

\newcommand{\prmark}{\textnormal{\textsc{conf}}}
\newcommand{\DC}{\textnormal{\textsc{DC}}}
\newcommand{\NDC}{\textnormal{\textsc{NDC}}}
\newcommand{\DCB}{\textnormal{\textsc{dcb-pr}}}
\newcommand{\NDCBConf}{\textnormal{\textsc{NDCB}}}

\newcommand{\norm}[1]{{\mathopen{\|}#1\mathclose{\|}}}

\begin{document}

\title[The Home-Space Problem for Petri Nets]{On the Home-Space Problem
for Petri Nets and its Ackermannian Complexity\rsuper*}
%\titlecomment{{\lsuper*}OPTIONAL comment concerning the title, \eg,
%  if a variant or an extended abstract of the paper has appeared elsewhere.}
\titlecomment{{\lsuper*}Combines and elaborates the results presented
at Concur'23 and at a workshop at ETAPS'24.}
%\thanks{thanks, optional.}	%optional

% affiliations are numbered automatically with a, b, c (see below)
% use the optional argument to indicate the affiliation(s) of each author
% omit the argument if there is only one author, or only one affiliation
\author[P.~Jan\v{c}ar]{Petr Jan\v{c}ar\lmcsorcid{0000-0002-8738-9850}}[a]
\author[J.~Leroux]{J\'er\^ome Leroux\lmcsorcid{0000-0002-7214-9467}}[b]

% affiliation 1 (automatically numbered a)
\address{Dept of Comp.\ Sci., Faculty of Science, Palack\'y Univ.\ Olomouc, Czechia} %optional
% write emails for all authors having that affiliation
\email{petr.jancar@upol.cz}  %optional

% affiliation 2 (automatically numbered b)
\address{LaBRI, CNRS, Univ. Bordeaux, France} %optional
\email{jerome.leroux@labri.fr}  %optional

%% etc.

%% required for running head on odd and even pages, use suitable
%% abbreviations in case of long titles and many authors:

%%%%%%%%%%%%%%%%%%%%%%%%%%%%%%%%%%%%%%%%%%%%%%%%%%%%%%%%%%%%%%%%%%%%%%%%%%%

%% the abstract has to PRECEDE the command \maketitle:
%% be sure not to issue the \maketitle command twice!

\begin{abstract}
  A set of configurations $H$ is a home-space for a set of configurations
  $X$ of a~Petri net if every configuration reachable from (any
  configuration in) $X$ can reach (some configuration in) $H$. The
  semilinear home-space problem for Petri nets asks, given a
  Petri net and semilinear sets of configurations $X$, $H$, if $H$ is
  a home-space for $X$. In 1989, David de Frutos Escrig and
  Colette Johnen proved that the problem is decidable when $X$ is
  a singleton and $H$ is a finite union of linear sets with the
  same periods. In this paper, we show that the general
  (semilinear) problem is decidable. This result is obtained by
  proving a duality between the reachability problem and the
  non-home-space problem. In particular, we prove that for any
  Petri net and any semilinear set of configurations $H$ we can
  effectively compute a~semilinear set $C$ of configurations,
  called a non-reachability core for $H$, such that for every set
  $X$ the set $H$ is not a home-space for $X$ if, and only if, $C$ is reachable from $X$. We show that the established relation to the reachability problem yields the Ackermann-completeness of the (semilinear) home-space problem. For this we also show that, given a Petri net with an initial marking, the set of minimal reachable markings can be constructed in Ackermannian time. 
\end{abstract}

\maketitle

%% start the paper here:

\section{Introduction}

On an abstract level, various practical systems and theoretical models
can be viewed as instances of 
 transition systems  $(S,\rightarrow)$ where $S$ is 
a~(possibly infinite) set of configurations 
and $\rightarrow\,\subseteq S\times S$ is a relation capturing when
one configuration
can change into another by an atomic step; 
the reachability relation $\reach$
is then the reflexive and transitive closure of $\rightarrow$.

Given a system  $(S,\rightarrow)$ and two sets $X,H\subseteq S$, we
say that $H$ is a \emph{home-space for} $X$ if from every
configuration reachable from (any configuration in) $X$ we can reach 
(some configuration in) $H$. 
The \emph{home-space problem} asks, given $(S,\rightarrow)$, $X$, $H$,
whether $H$ is a~home-space for $X$. For instance, 
the home-space problem can ask whether the system can always
return to an initial configuration. 
This paper focuses on the \emph{semilinear home-space problem for Petri
nets},
in which the respective sets $X,H$ are semilinear sets (consisting of
nonnegative integer vectors of a given dimension).

We recall that Petri nets provide a popular formal method
for modelling and analyzing parallel processes.
The standard model is not Turing-complete, and many analyzed
properties are decidable; we can refer to~\cite{survey-esparza} as to
one of the first survey papers on this issue.

A central algorithmic problem for Petri nets is reachability: given a
Petri net $A$ and two configurations $\vec{x}$ and $\vec{y}$, decide
whether there exists an execution of $A$ from $\vec{x}$ to $\vec{y}$.
In fact, many important computational problems in logic and complexity
reduce or are even equivalent to this problem (we can refer, e.g.,
to~\cite{DBLP:journals/siglog/Schmitz16,hack75} to exemplify this).
It was nontrivial to show that the reachability problem is
decidable~\cite{Mayr84},
and recently the complexity of this problem was
proved to be extremely high, namely
Ackermann-complete
(see~\cite{DBLP:conf/lics/LerouxS19} for the upper-bound
and~\cite{DBLP:journals/jacm/CzerwinskiLLLM21,DBLP:conf/focs/Leroux21,DBLP:conf/focs/CzerwinskiO21} for
the lower-bound).

The reachability problem for Petri nets can be generalized to
semilinear sets, a class of geometrical sets that coincides with the
sets definable in Presburger arithmetic~\cite{GS-PACIF66}. The
semilinear reachability problem for Petri nets asks, given a Petri net
$A$ and (presentations of) semilinear sets of configurations
$\vec{X}$,$\vec{Y}$, if there exists an execution from a configuration
in $\vec{X}$ to a configuration in $\vec{Y}$. Denoting by
$\post(\vec{X})$ the set of configurations reachable from $\vec{X}$
and by $\pre(\vec{Y})$ the set of configurations that can reach a
configuration in $\vec{Y}$, the semilinear reachability problem thus
asks, in fact,
if the intersection
$\post(\vec{X})\cap\pre(\vec{Y})$ is nonempty 
(which is equivalent to the non-emptiness of $X\cap\pre(\vec{Y})$ or $\post(\vec{X})\cap \vec{Y}$). 
This problem can be easily reduced to the classical reachability problem for Petri nets
 (where $\vec{X}$ and $\vec{Y}$ are singletons).

The semilinear home-space problem is a problem that seems to be
similar to the semilinear reachability problem at first sight. This
problem asks, given a Petri net $A$, and two semilinear sets
$\vec{X},\vec{H}$, if every configuration reachable from $\vec{X}$ can
reach $\vec{H}$, hence if $\post(\vec{X})\subseteq \pre(\vec{H})$. In
1989, David de Frutos Escrig and Colette Johnen~\cite{EscrigJohnen89}
proved that the semilinear home-space problem is decidable for
instances where $\vec{X}$ is a singleton set and $\vec{H}$ is a finite
union of linear sets using the same periods; they left the general case
open. 
In fact, the general problem seems close to the
decidability/undecidability border, since the reachability set
inclusion 
problem, which can be viewed as asking if
$\post(\vec{x})\subseteq\textsc{pre}^*_B(\vec{y})$ 
where $A,B$ are Petri nets of the same dimension (i.e., with the same
sets of places),
is 
undecidable~\cite{Baker73,DBLP:journals/tcs/Hack76},
even when
the dimension of $A,B$ is fixed to a~small value~\cite{DBLP:journals/tcs/Jancar95}.

\emph{Our contribution.}
In this paper, we show that the general semilinear home-space problem
is decidable. This result is obtained by proving a duality between the
reachability problem and the non-home-space problem. A crucial point
consists in proving that for any Petri net $A$ and for any linear set of
configurations $\vec{L}$, we can effectively compute a semilinear
``non-reachability core'' $\vec{C}$ 
such that for every set $\vec{X}$ the set $\vec{L}$ is not a
home-space for $\vec{X}$ if, and only if, $\vec{C}$ is reachable from $\vec{X}$.
By a technical analysis using the known complexity results for
reachability we show that 
the (semilinear) home-space problem is Ackermann-complete.
As an ingredient, we also show that, given a Petri net with an initial marking, the set of minimal reachable markings can be constructed in Ackermannian time.
Moreover, by using the results on 
inductive semilinear
invariants~\cite{DBLP:journals/corr/abs-1009-1076} we also show that a
semilinear, and moreover inductive, non-reachability core can be
computed for any semilinear (not only linear) set $\vec{S}$. This
yields a modification of the decidability proof but without complexity
bounds. We remark that only recently it has turned out that  Ackermannian upper bounds could be
derived in this way as well, due to the enhancement~\cite{DBLP:conf/fossacs/Leroux24} 
of~\cite{DBLP:journals/corr/abs-1009-1076}.
Finally, we also discuss the form of positive and negative witnesses
of the home-space property.

\emph{Organization of the paper.}
Section~\ref{sec:genapproach} describes
an idea of our approach in the context of general transition systems.
Section~\ref{sec:prelim} states our main results for
(transition systems generated by) Petri nets, after providing necessary preliminaries. 
Section~\ref{sec:HSPishard} shows the hardness results, yielding the
complexity lower bounds, and 
Sections~\ref{sec:nonhomespacewitness} and~\ref{sec:witnessfor linear}
give a decidability proof. Sections~\ref{sec:Ackone}
and~\ref{sec:Acktwo} contain the complexity analysis, yielding the
Ackermannian upper bounds.
Section~\ref{sec:semilcore} provides a proof that
any semilinear set admits an effectively computable
inductive semilinear non-reachability core.
In Section~\ref{sec:witnesses} we discuss the question
of positive and negative witnesses of the home-space property.
We conclude by a few remarks in Section~\ref{sec:conclusions}.

\section{A General Approach to the Home-Space Problem}\label{sec:genapproach}
In this section we provide an overview of the way the home-space problem 
can be solved via the so-called \emph{non-reachability cores}.
Though we apply this  approach to Petri nets, we start with presenting it
for a general transition system given as a pair $(S,\rightarrow)$ where $S$ is 
a~(possibly infinite) set of \emph{states} (or \emph{configurations})
and $\rightarrow\,\subseteq S\times S$ is 
a~\emph{transition relation}. 
The \emph{reachability relation} $\reach\,\subseteq S\times S$
is then the reflexive and transitive closure of $\rightarrow$.
For sets $X\subseteq S$, we introduce the following notions and
notation:
\begin{quote}
\begin{itemize}
	\item  by $\overline{X}$ we denote
the complement of $X$ (hence $S\smallsetminus X$);
\item
	$\prest(X)=\{s\in S\mid \exists s'\in X: s\reach s'\}$;
\item
	$\postst(X)=\{s\in S\mid \exists s'\in X: s'\reach s\}$;
\item $X$ is \emph{inductive} (or \emph{closed w.r.t. $\rightarrow$}) 
	if $\postst(X)=X$;
\item $H\subseteq S$ is a \emph{home-space} for $X$ if
	$\postst(X)\subseteq \prest(H)$.
\end{itemize}
\end{quote}
We might implicitly use simple observations like the following ones:
\begin{quote}
\begin{itemize}
	\item
		$X\subseteq\prest(X)=\prest(\prest(X))$,
\item
$\prest(X_1\cup X_2)=\prest(X_1)\cup \prest(X_2)$, 
\item
if both $X_1$ and $X_2$ are inductive, then $X_1\cap X_2$ is
		inductive.
\end{itemize}
\end{quote}
	We also observe that $H$ is a home-space for $X$ if{f} 
it is a~home-space for every $s\in X$ (implicitly viewed as the singleton
$\{s\}$).

\begin{figure}[t]
\centering{\includegraphics[width=0.8\textwidth]{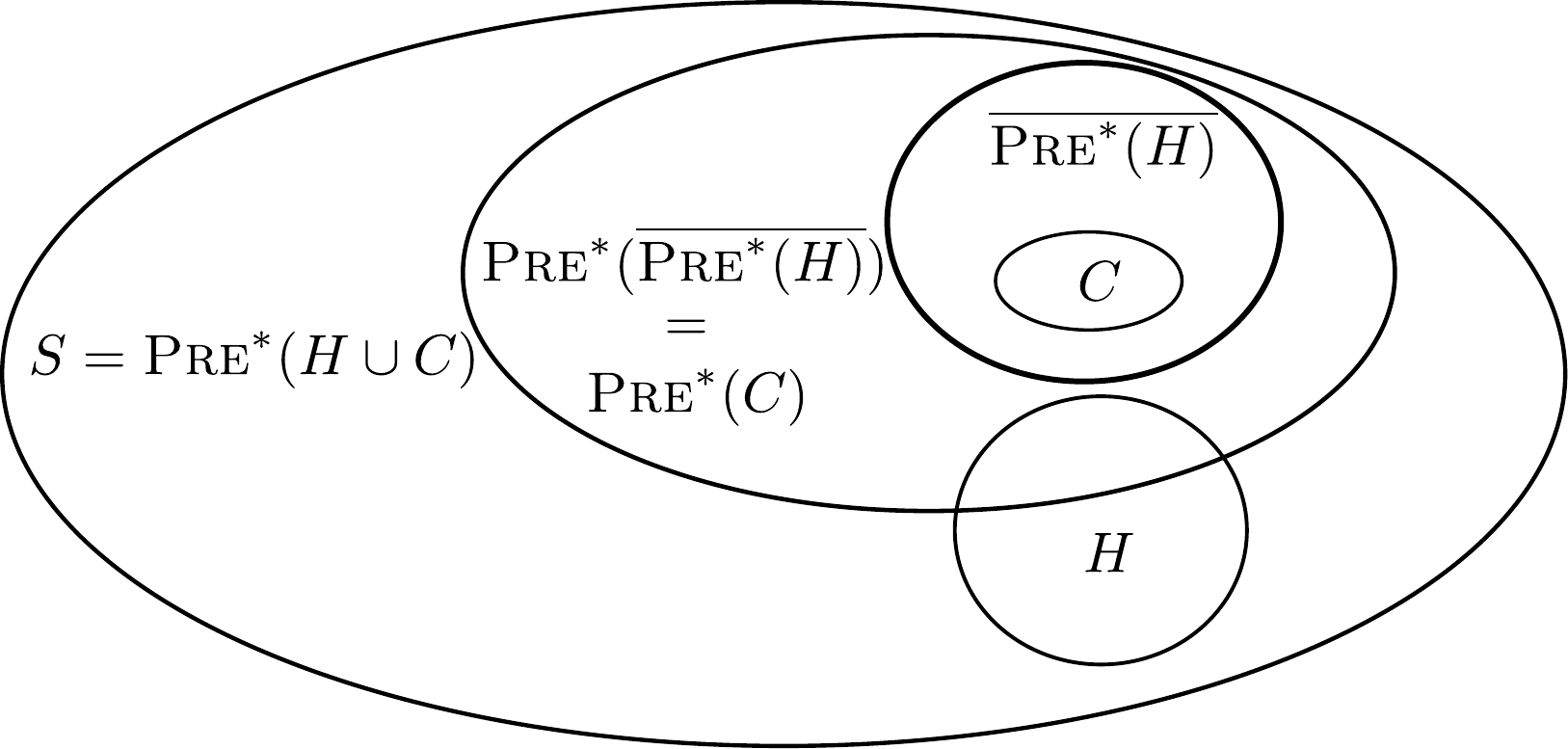}}
\caption{$C$ is a non-reachability core for
	$H$.}\label{fig:nonreachcore}
\end{figure}

Figure~\ref{fig:nonreachcore} depicts the set $S$ of states of a~system, and a subset
$H\subseteq S$ as a~potential ``home-space'' in which we are interested.
The set $\overline{\prest(H)}$ consists of the states from
which $H$ is not reachable, which entails that $\overline{\prest(H)}$
is inductive (i.e., $\postst(\overline{\prest(H)})=\overline{\prest(H)}$).
We observe that
\begin{center}
 $\prest(\overline{\prest(H)})=\{s\in S\mid H$ is not a home-space for
	$s\}$;
\end{center}
hence $H$ is a home-space for $X$ if{f} $X\cap
\prest(\overline{\prest(H)})=\emptyset$.

We also note that $H$ and $\overline{\prest(H)}$ are disjoint, but
$\prest(\overline{\prest(H)})$ might intersect $H$.
Since $S=\prest(H)\cup\overline{\prest(H)}$, we have
\begin{center}
$S=\prest(H\cup
\overline{\prest(H)})$.
\end{center}

\paragraph{Non-Reachability Cores}

For some (infinite-state) systems it might be hard to construct 
(a description of) the set  
 $\overline{\prest(H)}$ and/or to decide for $s\in S$ whether
 $s\in \prest(\overline{\prest(H)})$.
Surely, for Turing-powerful systems such problems
 are not algorithmically solvable. But in the case
 of Petri nets it has turned out useful to introduce the notion
of a \emph{non-reachability core}, or just a \emph{core}, \emph{for}
$H$:
it is a set $C\subseteq S$
 (also depicted in Figure~\ref{fig:nonreachcore}) such that
\begin{center}
	$C\subseteq\overline{\prest(H)}\subseteq\prest(C)$, 
\end{center}
which
	entails that 
	$S=\prest(H\cup C)$ (since $S=\prest(H)\cup\overline{\prest(H)}$);
in other words, $C$ is a subset of 
$\overline{\prest(H)}$ 
that is its home-space (i.e., $C$ is a home-space for $\overline{\prest(H)}$). 
Hence if $C$ is a core for $H$, then 
	\[\prest(\overline{\prest(H)})=\prest(C);\]
	therefore $H$ is not a home-space for $X$ if{f} $C$ is
	reachable from some $s\in X$.

Of course, such a notion can help us only if
there are cores $C$ for $H$ that are somehow simpler than $\overline{\prest(H)}$ itself.
We have noted that $\overline{\prest(H)}$ is inductive; the cores
$C\subseteq\overline{\prest(H)}$ do not
need to be inductive, but inductive non-reachability cores will be of
special interest for us.

\paragraph{Non-Reachability Cores in Finite-State Systems}

\begin{figure}[t]
\usetikzlibrary {arrows.meta,automata,positioning} 
\begin{tikzpicture}[shorten >=1pt,node distance=2cm,on grid,>={Stealth[round]},
    every state/.style={draw=black,very thick}]

  \node[state]  (s_0)                      {$s_0$};
  \node[state]          (s_1) [above right=of s_0] {$s_1$};
  \node[state]          (s_2) [below right=of s_0] {$s_2$};
  \node[state] (s_3) [above right=of s_2] {$s_3$};
  \node[state] (s_4) [below right=of s_3] {$s_4$};
  \node[state] (s_5) [right=of s_3] {$s_5$};
  \node[state] (s_6) [right=of s_5] {$s_6$};
  \node[state] (s_7) [right=of s_6] {$s_7$};
  \node[state] (s_8) [right=of s_7] {$s_8$};
  \node[state] (s_9) [above right=of s_3] {$s_9$};
  \node[state] (s_10) [right=of s_9] {$s_{10}$};

  \path[->] (s_0) edge (s_2);
  \path[->] (s_1) edge (s_0);
  \path[->] (s_2) edge (s_3);
  \path[->] (s_3) edge (s_1);
  \path[->] (s_3) edge (s_4) edge (s_9) edge (s_5);
  \path[->] (s_6) edge (s_7);
  \path[->] (s_5) edge [bend left] (s_6);
  \path[->] (s_6) edge [bend left] (s_5);
  \path[->] (s_7) edge [bend left] (s_8);
  \path[->] (s_8) edge [bend left] (s_7);
  \path[->] (s_9) edge [bend left] (s_10);
  \path[->] (s_10) edge [bend left] (s_9);
  
\end{tikzpicture}
\caption{Let $H=\{s_3,s_4\}$. We have $\prest(H)=\{s_0,s_1,s_2,s_3,s_4\}$,
	and the bottom SCCs of $\overline{\prest(H)}$ are $\{s_7,s_8\}$ and $\{s_9,s_{10}\}$. 
	Hence $C=\{s_7,s_9\}$ is one non-reachability core for
	$H$.
}\label{fig:example}
\end{figure}
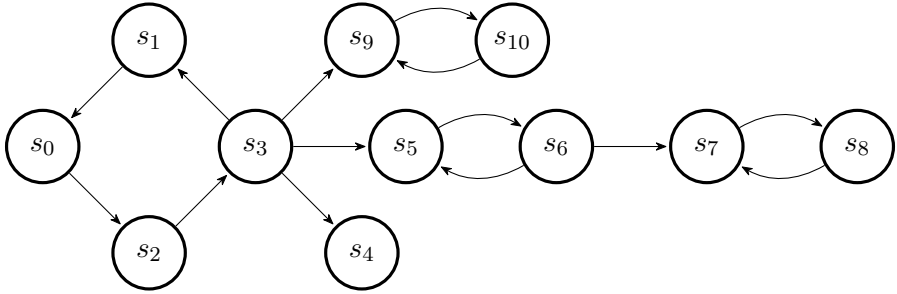

It is straightforward to characterize the non-reachability cores in
finite-state systems, which are exemplified by the system in
Figure~\ref{fig:example}. We can partition
$\overline{\prest(H)}$ into the strongly connected components (SCCs),
and observe that a~set $C$ is a~core if, and only if, it is included in  
$\overline{\prest(H)}$ and contains at least one state in each bottom
SCC of $\overline{\prest(H)}$
(from which no other SCC is reachable).

\paragraph{Non-Reachability Cores for Unions of Sets}

A ``home-space'' set $H\subseteq S$ can be sometimes naturally given
as the union of smaller sets 
(in the case of Petri nets we are interested in semilinear home-space sets,
which are defined as finite unions of linear sets).
For instance, in Figure~\ref{fig:example} we have $H=H_1\cup H_2$ where
$H_1=\{s_3\}$ and $H_2=\{s_4\}$.
We can consider $C_1=\{s_9,s_7,s_4\}$ as a core for $H_1$ and $C_2=\{s_{10},s_8\}$
as a core for $H_2$.

Having some cores $C_1,C_2,\dots,C_m$
for sets $H_1,H_2,\dots,H_m$, respectively, 
it is natural to ask if
we can combine these cores to get a core $C$ for the
set  $H=H_1\cup H_2\cdots\cup H_m$. 
Proposition~\ref{prop:union} gives a simple answer if the cores $C_i$
are inductive: then the intersection of the cores $C_i$ is such a core
$C$, which is, moreover, inductive.
It will turn out that this fact is sufficient for developing a
decidability proof for the semilinear home-space problem for Petri
nets; in particular we will show that any semilinear set has an
effectively constructible semilinear non-reachability core that is
inductive.
Nevertheless,
 for deriving the complexity upper bound we will use
Proposition~\ref{prop:coreexecution} that does not require the cores
to be inductive. 

\begin{prop}\label{prop:coreexecution} 
  Given $(S,\rightarrow)$ and $H\subseteq S$, let $H=H_1\cup
  H_2\cdots\cup H_m$ for some $m\geq 1$, and let
  $C_1,C_2,\dots,C_m$ be non-reachability cores for
  $H_1,H_2,\dots,H_m$, respectively.
  For each $X\subseteq S$ we have that
  $H$ is not a~home-space for $X$ if, and only if, there is an execution
  \begin{equation}\label{eq:badcomput}
    s_0\reach s_1\reach s_2\cdots\reach
    s_m
  \end{equation}
  where $s_0\in X$, and $s_1\in C_1$, $s_2\in C_2$, $\ldots$, $s_m\in
	C_m$.
\end{prop}
\begin{proof}
  Given an execution~(\ref{eq:badcomput}), the facts 
  that $s_i\in C_i$ and $C_i$ is a
  non-reachability core for
  $H_i$ (hence $C_i\subseteq\overline{\prest(H_i)}$) entail $s_i\not\reach H_i$, for all
	$i\in\{1,2,\dots,m\}$.
  The facts $s_i\not\reach H_i$ and $s_i\reach s_m$
  entail that $s_m\not\reach H_i$ (for all
	$i\in\{1,2,\dots,m\}$).
  Hence $s_m\not\reach H$
  (where $H=H_1\cup H_2\cdots\cup H_m$),
  and the facts
  $s_0\in X$ and
  $s_0\reach s_m\not\reach H$ entail that
  $H$ is not a~home-space for $X$.
  
  Conversely, we consider a set  $X\subseteq S$ for which
  $H$ is not a home-space.
  Hence there exist configurations $s_0,s'_0$ such that
  $s_0\in X$ and $s_0\reach s'_0\not\reach H$. 
  In particular $s'_0\not\reach H_1$, and thus
  $H_1$ is not a home-space for $s'_0$.
  Since $C_1$ is a non-reachability core for $H_1$, we have 
  $s'_0\reach s_1$ for some $s_1\in C_1$.
  Since $s'_0\not\reach H$ and $s'_0\reach s_1$,
  we have $s_1\not\reach H$, and in particular 
  $s_1\not\reach H_2$. Since $H_2$ 
  is not a home-space for $s_1$ and $C_2$ is a
  non-reachability core for $H_2$, we get  $s_1\reach s_2$
  for some $s_2\in C_2$. Continuing in this way, we successively
  derive the existence of an execution~(\ref{eq:badcomput}).
\end{proof}

\begin{prop}\label{prop:union}
Given $(S,\rightarrow)$ and $H\subseteq S$, let $H=H_1\cup
  H_2\cdots\cup H_m$ for some $m\geq 1$, and let
  $C_1,C_2,\dots,C_m$ be inductive
  non-reachability cores for
  $H_1,H_2,\dots,H_m$, respectively.
  Then $C_1\cap C_2\cdots\cap C_m$ is an inductive non-reachability core for $H$.
\end{prop}
\begin{proof}
  By induction on $m$.
  The case $m=1$ is trivial, 
  so we now suppose $m=2$, hence $H=H_1\cup H_2$. Since $C_1$ and $C_2$ are inductive, the
  intersection $C=C_1\cap C_2$ is inductive as well.
  Since $C_1\subseteq\overline{\prest(H_1)}$ and
  $C_2\subseteq\overline{\prest(H_2)}$, we have 
  \[C_1\cap C_2\subseteq \overline{\prest(H_1)}\cap\overline{\prest(H_2)}=
  \overline{\prest(H_1)\cup\prest(H_2)}=\overline{\prest(H_1\cup
    H_2)},\]
  hence $C\subseteq \overline{\prest(H)}$.
  
  Let us show that 
  $\overline{\prest(H)}\subseteq\prest(C)$; 
  we recall that 
  $\overline{\prest(H)}$ is inductive.
  If $s\in\overline{\prest(H)}$, 
  then
  \[\postst(\{s\})\subseteq\overline{\prest(H)}=
  \overline{\prest(H_1\cup H_2)}=
  \overline{\prest(H_1)}\cap\overline{\prest(H_2)}.\]
  Since $s\in\overline{\prest(H_1)}$, there is $s_1\in C_1$ such that
  $s\reach s_1$. Using the fact that $C_1$ is inductive, we
  deduce that 
  \[\postst(\{s_1\})\subseteq C_1\cap\overline{\prest(H_1)}\cap\overline{\prest(H_2)}.\]
  Since $s_1\in\overline{\prest(H_2)}$, there is $s_2\in C_2$ such that
  $s_1\reach s_2$; we thus have $s_2\in C_1\cap C_2$. Since
  $s\reach s_2$, we have shown that
  $\overline{\prest(H)}\subseteq\prest(C)$, 
  which finishes the
  proof that $C=C_1\cap C_2$ is a non-reachability core for
  $H=H_1\cup H_2$.
  
  The claim for $m\geq 3$ follows by the induction hypothesis,
  since $H_1\cup H_2\cdots\cup H_m$ can be viewed as
  $H_1\cup H_2\cdots\cup H_{m-2}\cup (H_{m-1}\cup H_m)$ where we
  consider $C_{m-1}\cap C_m$ as 
  the inductive core for the set $(H_{m-1}\cup H_m)$.
\end{proof}

\section{Basic Notions, and Main Results}\label{sec:prelim}

In this section we state the main results,
which deal with 
 transitions systems $(S,\rightarrow)$
generated by (unmarked place/transition) Petri nets.
We start with
introducing basic notions and
notation.

By $\setN$ we denote the set $\{0,1,2,\dots\}$ of nonnegative
integers. For $i,j\in\setN$, by $[i,j]$ we denote the set $\{i,i{+}1,\dots,j\}$
(which is empty if $i>j$).

\paragraph{Notation for Vectors of Nonnegative Integers}

For (a dimension) $d\in\setN$, the elements of $\setN^d$ are called
($d$-dimensional) \emph{vectors}; they are
denoted in bold face, and for $\vec{x}\in\setN^d$ we write 
\begin{center}
$\vec{x}=(\vec{x}(1),\vec{x}(2),\ldots,\vec{x}(d))$ 
\end{center}
	so that we can
refer to the vector components.
We use the component-wise sum $\vec{x}+\vec{y}$ of vectors, and
their component-wise order  $\vec{x}\leq\vec{y}$. For $c\in\setN$, we write
\begin{center}
$c\cdot\vec{x}=(c\cdot\vec{x}(1),c\cdot\vec{x}(2),\ldots,c\cdot\vec{x}(d))$.
\end{center}
By the \emph{norm of} $\vec{x}$, denoted  $\|\vec{x}\|$, we mean the sum of components, i.e.,
$\|\vec{x}\|=\sum_{i=1}^d \vec{x}(i)$. 

By $\vec{0}$ we denote the zero vector whose dimension is always clear
from its context. Occasionally we slightly abuse notation by
presenting a vector as a mix of 
subvectors and integers; in particular,
given $\vec{x}\in\setN^d$ and $y_1,y_2,\dots,y_m\in \setN$, 
we might write $(\vec{x},y_1,y_2,\dots,y_m)$ to denote the
$(d{+}m)$-dimensional vector 
$(\vec{x}(1),\vec{x}(2),\ldots,\vec{x}(d),y_1,y_2,\dots,y_m)$.

Given a set $\vec{X}\subseteq\setN^d$, by $\overline{\vec{X}}$ we denote its
complement, i.e., $\overline{\vec{X}}=\setN^d\smallsetminus \vec{X}$.

\paragraph{Linear and Semilinear Sets of Vectors, and their
Presentations}

A \emph{set} $\vec{L}\subseteq \setN^d$ is \emph{linear}
if there are $d$-dimensional vectors $\vec{b}$, the \emph{basis}, and
$\vec{p}_1,\vec{p}_2,\ldots,\vec{p}_k$, the \emph{periods} (for
$k\in\setN$), such that 
\begin{center}
$\vec{L}=\{\vec{x}\in\setN^d\mid
	\vec{x}=\vec{b}+\vec{u}(1)\cdot\vec{p}_1+\vec{u}(2)\cdot\vec{p}_2
		\cdots
		+\vec{u}(k)\cdot\vec{p}_k$ for some
		$\vec{u}\in\setN^k\}$.
\end{center}
	In this case, by a \emph{presentation of} $\vec{L}$ we mean the tuple
$(\vec{b},\vec{p}_1,\vec{p}_2,\ldots,\vec{p}_k)$.

A \emph{set} $\vec{S}\subseteq \setN^d$ is \emph{semilinear} if it is a
finite union of linear sets, i.e. 
\begin{center}
$\vec{S}=\vec{L}_1\cup\vec{L}_2\cdots \cup \vec{L}_m$ 
\end{center}
	where 
$\vec{L}_i$ are linear sets (for all $i\in[1,m]$). 
In this case, by a \emph{presentation of} $\vec{S}$ we mean the sequence 
of presentations of $\vec{L}_1,\vec{L}_2,\ldots,\vec{L}_m$.
When we say that a \emph{semilinear set} $\vec{S}$ is given, we mean that we are
given a presentation of $\vec{S}$; when we say that $\vec{S}$
is \emph{effectively constructible} in some context, we mean that there
is an algorithm computing its presentation (in the respective context). 

\paragraph{Semilinear sets and Presburger arithmetic}
We recall that a set $\vec{S}\subseteq\setN^d$ is semilinear
if, and only if, it is expressible in Presburger arithmetic~\cite{GS-PACIF66};
the respective transformations between presentations and formulas are
effective. 
Hence if  $\vec{S}\subseteq\setN^d$ is semilinear, then also its
complement $\overline{\vec{S}}$ is semilinear, and
$\overline{\vec{S}}$ is effectively constructible when (a presentation
of) $\vec{S}$ is given.

\paragraph{Petri Nets}

We use a concise definition of (unmarked place/transition)
Petri nets. By a~\emph{$d$-dimensional Petri-net action} we mean  
a pair $a=(\vec{a}_-,\vec{a}_+)\in\setN^d\times\setN^d$.
With  $a=(\vec{a}_-,\vec{a}_+)$  we associate the binary relation $\xrightarrow{a}$
on the set $\setN^d$ by putting
$(\vec{x}+\vec{a}_-)\xrightarrow{a}(\vec{x}+\vec{a}_+)$ for all
$\vec{x}\in\setN^d$. The relations $\xrightarrow{a}$ are naturally
extended to the relations $\xrightarrow{\sigma}$ for finite sequences
$\sigma$ of ($d$-dimensional Petri net) actions.

A \emph{Petri net} $A$ \emph{of dimension} $d$ (with $d$
places in more traditional definitions) is a~finite
set of $d$-dimensional Petri-net actions (transitions).
Here the vectors $\vec{x}\in\setN^d$ are also called
\emph{configurations} (markings).
On the set $\setN^d$ of configurations we define the
\emph{reachability relation} that we now denote by 
$\reachA$ (instead of $\reach$), to highlight the underlying Petri net $A$: we write
 $\vec{x}\reachA\vec{y}$ if there is $\sigma\in A^*$ 
such that $\vec{x}\xrightarrow{\sigma}\vec{y}$. 
For $\vec{x}\in\setN^d$ and $\vec{X}\subseteq \setN^d$ we put
\begin{center}
$\post(\vec{x})=\{\vec{y}\in\setN^d\mid \vec{x}\reachA \vec{y}\}$, and
$\post(\vec{X})=\bigcup_{\vec{x}\in
\vec{X}} \post(\vec{x})$.
\end{center}
Symmetrically, for $\vec{y}\in\setN^d$ and $\vec{Y}\subseteq \setN^d$
we put
\begin{center}
$\pre(\vec{y})=\{\vec{x}\in\setN^d\mid \vec{x}\reachA \vec{y}\}$
and
$\pre(\vec{Y})=\bigcup_{\vec{y}\in \vec{Y}} \pre(\vec{y})$. 
\end{center}
By $\vec{X}\reachA \vec{Y}$ we denote that
$\vec{x}\reachA\vec{y}$ for some $\vec{x}\in\vec{X}$ and
$\vec{y}\in\vec{Y}$, i.e.\ that $\post(\vec{X})\cap
\vec{Y}\neq\emptyset$, or equivalently $\vec{X}\cap
\pre(\vec{Y})\neq\emptyset$.

\paragraph{(Semilinear) Reachability Problem}

By the (semilinear) \emph{reachability problem} we mean the following
decision problem:
\begin{quote}
	\emph{Instance:} a $d$-dimensional Petri net $A$ and
	presentations of two semilinear sets
	$\vec{X},\vec{Y}\subseteq\setN^d$, which we refer
	to as the triple $A,\vec{X},\vec{Y}$.
\\
	\emph{Question:} does $\vec{X}\reachA \vec{Y}$ 
	hold?
\end{quote}
In the standard definition of the reachability problem the sets
$\vec{X},\vec{Y}$ are singletons; the problem is
decidable~\cite{Mayr84}, and it has been recently shown to be 
Ackermann-complete~\cite{DBLP:conf/lics/LerouxS19,DBLP:conf/focs/Leroux21,DBLP:conf/focs/CzerwinskiO21}.
It is well-known (and easy to
show) that the above more general version (the semilinear reachability
problem) is tightly related to
the standard version, and has thus the same complexity.

\begin{rem}
We can sketch this tight relation as follows.
	If $\vec{X}$ and $\vec{Y}$ are linear,
with presentations $(\vec{b},\vec{p}_1,\vec{p}_2,\dots,\vec{p}_{k})$
and $(\vec{b}',\vec{p}'_1,\vec{p}'_2,\dots,\vec{p}'_{k'})$
	respectively, then it suffices to ask whether
	$\vec{b}\xrightarrow{(A')^*} \vec{b}'$ where $A'$ arises from
	$A$ by adding the actions $(\vec{0},\vec{p}_i)$ for all
	$i\in[1,k]$ and $(\vec{p}'_i,\vec{0})$ for all $i\in[1,k']$.
	Now if $\vec{X}=\vec{L}_1\cup\vec{L}_2\cdots\cup \vec{L}_m$
	and  $\vec{Y}=\vec{L}'_1\cup\vec{L}'_2\cdots\cup
	\vec{L}'_{m'}$, then it suffices to check if $\vec{L}_i\reachA
	\vec{L}'_j$ for some $i\in[1,m]$ and $j\in[1,m']$.
(In fact, there is also a polynomial 
	reduction of the general version
	to the standard one, which increases the dimension.)	
\end{rem}

\paragraph{Semilinear Home-Space Problem}

For a Petri net $A$ of dimension $d$ and two 
sets $\vec{X},\vec{H}\subseteq\setN^d$, by following
the definitions introduced in the previous section
we call $\vec{H}$ 
a~\emph{home-space for} $(A,\vec{X})$, or just
for 
$\vec{X}$ when $A$ is clear from the context, if $\post(\vec{X})\subseteq \pre(\vec{H})$. 
We note that the above
(semilinear) reachability problem in fact asks, given
$A,\vec{X},\vec{Y}$, if $\post(\vec{X})\cap
\pre(\vec{Y})\neq\emptyset$. 
The \emph{semilinear home-space problem} is defined as follows:
\begin{quote}
	\emph{Instance:} a triple $A,\vec{X},\vec{H}$ where $A$ is a
	Petri net, of dimension $d$, and $\vec{X}$, $\vec{H}$ are two
	(finitely presented)
	semilinear subsets of $\setN^d$.
\\
	\emph{Question:} is
	$\post(\vec{X})\subseteq
	\pre(\vec{H})$ (i.e., is $\vec{H}$ a home-space for $\vec{X}$)~?
\end{quote}
\paragraph{Main Results}
Our main result is stated by Theorem~\ref{th:AckerComplex}.
Nevertheless, we first prove the weaker claim,
Theorem~\ref{th:semhomspace}, that
answers an open
question from~\cite{EscrigJohnen89} and does not need the
technicalities related to the complexity analysis. 

\begin{thm}\label{th:semhomspace}
The semilinear home-space problem is decidable.
\end{thm}

\begin{thm}\label{th:AckerComplex}
The semilinear home-space problem is Ackermann-complete.
\end{thm}
\noindent %FIXED indent
We remark that by~\cite{EscrigJohnen89} we know that the home-space problem  
is decidable for the instances $A$, $\vec{X}$, $\vec{H}$ where $\vec{X}$ is a singleton set, 
and $\vec{H}$ is a finite union of linear sets with the same periods;
this was established by a Turing reduction to the reachability problem.
The decidability in the case where $\vec{H}$ is a general semilinear set 
was left open in~\cite{EscrigJohnen89}; this more general problem
indeed looks more subtle but we manage to provide a solution here.
Before doing this, we note in Section~\ref{sec:HSPishard} that the
problem has also a high computational complexity, and can be naturally
viewed as residing at the decidability/undecidability border.

\begin{rem}
Some intermediate results that help us to derive Theorems~\ref{th:semhomspace}
	and~\ref{th:AckerComplex} seem to be interesting on
	their own. In particular we name Theorem~\ref{thm:core}
	showing that for each semilinear set we can effectively construct its
	inductive semilinear non-reachability core.
Another example is an Ackermannian-time algorithm constructing the
	minimal elements in the reachability set of a given Petri net
	with an initial configuration (i.e., in the set
	$\post(\vec{x})$), which is given in 
Section~\ref{sec:minreachconfig}.
\end{rem}

\section{The Home-Space Problem is Hard}\label{sec:HSPishard}
We first note that even a simple version of the home-space problem is
at least as hard as (non)reachability, and thus Ackermann-hard.
We use a polynomial reduction that increases the Petri net dimension,
by additional
vector components that can be viewed as control states. (It would be natural to
use the model of \emph{vector addition systems with states}
but we do not introduce them formally in this paper.)

\begin{prop}\label{prop:AckLowerBound}
  The non-reachability problem is polynomially 
  reducible to the home-space problem restricted to
  the instances $A,\vec{X},\vec{H}$ where $\vec{X}$ and $\vec{H}$ are singletons.
\end{prop}
\begin{proof}
  Let us consider a~Petri net $A$ of dimension $d$ and
  two vectors $\vec{x},\vec{y}\in\setN^d$, as an
  instance of the (non)reachability problem.
  We create the $(d{+}3)$-dimensional Petri net $A'$ so
  that each action $a=(\vec{a}_-,\vec{a}_+)$ of $A$ is
  transformed to the action
  $a'=((\vec{a}_-,1,0,0),(\vec{a}_+,1,0,0))$ of $A'$, and
  $A'$ has also the additional actions
  $((\vec{y},1,0,0),(\vec{0},0,1,0))$,
  $((\vec{0},1,0,0),(\vec{0},0,0,1))$,  
  and the actions $((\vec{i}_j,0,1,0),(\vec{0},0,0,1))$,
  $((\vec{i}_j,0,0,1),(\vec{0},0,0,1))$ for all
  $j\in[1,d]$, where $\vec{i}_j\in\setN^d$ satisfies
  $\vec{i}_j(j)=1$ and  $\vec{i}_j(i)=0$
  for all $i\neq j$.
  
	We verify that $\vec{x}\reachA \vec{y}$ if{f} $\{(\vec{0},0,0,1)\}$ is not a home-space for
  $(A',\{(\vec{x},1,0,0)\})$:
  \begin{itemize}
  \item
    if $\vec{x}\reachA \vec{y}$, then
    $(\vec{x},1,0,0)\xrightarrow{(A')^*}
    (\vec{y},1,0,0)\xrightarrow{(A')^*}(\vec{0},0,1,0)$,
    and $(\vec{0},0,0,1)$ is not reachable
    from $(\vec{0},0,1,0)$;
  \item
    if $\vec{x}\not\reachA \vec{y}$, then any configuration reachable from 
    $(\vec{x},1,0,0)$ in $A'$ is in one of
    the forms $(\vec{y}',1,0,0)$,
    $(\vec{z},0,1,0)$, $(\vec{z}',0,0,1)$
    where $\vec{y'}\neq\vec{y}$ and
    $\vec{z}\neq\vec{0}$, and
    $(\vec{0},0,0,1)$ is clearly reachable
    from all of them.
    \qedhere
  \end{itemize}
\end{proof}

\noindent %FIXED indent
Now we note that a slight generalization of the semilinear home-space
problem is undecidable; it is the case when instead of semilinear
sets $\vec{H}$ in the instances $A,\vec{X},\vec{H}$ we allow
$\vec{H}$ to be reachability sets of Petri nets 
(that are a special case of so called \emph{almost semilinear
sets}~\cite{DBLP:conf/birthday/Leroux12}). 
	
\begin{prop}\label{prop:undecinclusion}
  Given Petri nets $A,B$ of the same dimension $d$, and two vectors
  $\vec{x},\vec{y}\in\setN^d$, it is undecidable if
  $\textsc{post}^*_B(\vec{y})$ is a home-space for
  $(A,\{\vec{x}\})$.
\end{prop}
\begin{proof}
  We recall that the reachability set inclusion problem is undecidable
  for Petri nets (and for the equivalent model of vector addition
  systems);
  see~\cite{Baker73,DBLP:journals/tcs/Hack76,DBLP:journals/tcs/Jancar95}.
  Hence it is undecidable, given Petri nets $A,B$ of the same dimension
  $d$ and $\vec{x},\vec{y}\in\setN^d$, whether
  $\post(\vec{x})\subseteq\textsc{post}^*_B(\vec{y})$.
  If $A'$ arises from $A$ by replacing each action
  $a=(\vec{a}_-,\vec{a}_+)$ with
  $a'=((\vec{a}_-,1),(\vec{a}_+,1))$ and by adding 
  the action $((\vec{0},1),(\vec{0},0))$,
  and $B'$ arises from $B$ by replacing each
  $b=(\vec{b}_-,\vec{b}_+)$ with
  $b'=((\vec{b}_-,0),(\vec{b}_+,0))$, then we obviously
  have that $\textsc{post}^*_{B'}((\vec{y},0))$ is 
  a~home-space for $(A',(\vec{x},1))$ if, and only if, 
  $\post(\vec{x})\subseteq\textsc{post}^*_B(\vec{y})$.
\end{proof}	

\begin{rem}\label{rem:fixedfivedim}
  Since~\cite{DBLP:journals/tcs/Jancar95} shows, in fact, that the 
  reachability set inclusion (or equality) problem is undecidable 
  even for some fixed five-dimensional
  vector addition
  systems with states (VASSs), we could appropriately strengthen
  Proposition~\ref{prop:undecinclusion}; but we do not pursue 
  this technical issue here.
\end{rem}

We can note that the undecidability of the question whether 
$\textsc{post}_B^*(\vec{x})\subseteq\textsc{post}^*_A(\vec{y})$ entails 
that
the question whether
$\textsc{post}_B^*(\vec{x}) \subseteq \textsc{pre}^*_A(\vec{y})$
is also undecidable
(since $\textsc{post}_A^*(\vec{y})$ is equal to
$\textsc{pre}_{A_{rev}}^*(\vec{y})$ where $A_{rev}$ arises from $A$ by
reversing each action  $(\vec{a}_-,\vec{a}_+)$ to
$(\vec{a}_+,\vec{a}_-)$).
On the other hand, in the next sections we show that the question whether 
$\post(\vec{x}) \subseteq \pre(\vec{y})$ is decidable.
We will show that, given a $d$-dimensional Petri net $A$ and
$\vec{y}\in\setN^d$, we can
effectively construct a semilinear non-reachability core $C\subseteq
\setN^d$ for $\{y\}$, where 
$\post(\vec{x}) \not\subseteq \pre(\vec{y})$ if, and only if, $\post(\vec{x})$
intersects $C$. The equality of the nets on both sides is crucial,
since if $\textsc{post}_B^*(\vec{x})$ does not intersect $C$, then
this does not entail 
$\textsc{post}_B^*(\vec{x}) \subseteq \pre(\vec{y})$.

\section{Decidability of Home-Space via Semilinear Non-Reachability Cores}\label{sec:nonhomespacewitness}
Now we start to discuss
how to decide the semilinear home-space problem.
We consider a~fixed Petri net $A$ of dimension $d$ if not said otherwise.

Since a semilinear set is a finite union of linear sets,
Proposition~\ref{prop:coreexecution} shows that the semilinear
home-space problem can be reduced to a form of the reachability
problem as soon as semilinear non-reachability cores can be computed
for linear sets:

\begin{lem}\label{lem:coreforlinear}
  Given a Petri net $A$ of dimension $d$, and (a presentation
  of) a linear set
  $\vec{L}\subseteq \setN^d$, there is an effectively constructible 
  semilinear non-reachability core $\vec{C}$ for $\vec{L}$.
\end{lem}
\noindent %FIXED indent
This crucial lemma will be proved in the next section
(Section~\ref{sec:witnessfor linear}).
Here we show the decidability of the semilinear home-space
problem when assuming the lemma. We note that the semilinear non-reachability
core claimed by the lemma is not necessarily inductive; that's why we
use  Proposition~\ref{prop:coreexecution}, and not
 Proposition~\ref{prop:union}.

The next proposition (related to  Proposition~\ref{prop:coreexecution}) gives us the final ingredient for showing an
algorithm deciding the semilinear home-space problem.
\begin{prop}\label{prop:decsemilinseq}
  Given a Petri net $A$ of dimension $d$, and (presentations of)
  semilinear subsets 
  $\vec{X}_0,\vec{X}_1,\ldots, \vec{X}_m$ of $\setN^d$,
  the existence of 
  an execution
  \begin{equation}\label{eq:stepexecution}
    \vec{x}_0\reachA \vec{x}_1\reachA \vec{x}_2\cdots\reachA
    \vec{x}_m
  \end{equation}
  where $\vec{x}_i\in \vec{X}_i$ for each
  $i\in[0,m]$
  is decidable (by a reduction
  to reachability).
\end{prop}
\begin{proof}
  By a standard construction, we can build a Petri net with a bigger
  dimension and an initial configuration that first generates $m$ copies of some
  $\vec{x}_0\in\vec{X}_0$, then performs an execution of $A$
  from $\vec{x}_0$ on all these copies, while  
  at some moment it freezes some configuration $\vec{x}_1$ reached in the first copy,
  later it freezes some $\vec{x}_2$ reached in the second copy, etc.; at the
  end it starts a
  ``testing part'' that enables to reach the zero configuration
  if, and only if,
  $\vec{x_1}\in\vec{X}_1$, $\vec{x}_2\in\vec{X}_2$, $\ldots$,
  $\vec{x}_m\in\vec{X}_m$.
\end{proof}

\noindent %FIXED indent
We note that 
a proof of Theorem~\ref{th:semhomspace} is now clear:
Given a Petri net $A$ of dimension $d$ and two semilinear sets 
$\vec{X}, \vec{H}\subseteq \setN^d$,
we use that $\vec{H}=\vec{H}_1\cup\vec{H}_2\ldots\cup \vec{H}_m$ where $\vec{H}_i$ are linear sets,
and by Lemma~\ref{lem:coreforlinear} we can construct a semilinear
non-reachability core $\vec{C}_i$ for
$\vec{H}_i$, for each $i\in[1,m]$. Then we ask if there is
an execution~(\ref{eq:badcomput}) from
Proposition~\ref{prop:coreexecution};
this can be decided effectively by
Proposition~\ref{prop:decsemilinseq}.

\section{Effective Semilinear Non-Reachability Core for Linear Set}\label{sec:witnessfor linear}

In Section~\ref{sec:MinElem}
we recall an important ingredient dealing with computing the minimal elements in
some set $\vec{X}\subseteq \setN^d$;  
its use in Petri nets originates in the
work by Valk and Jantzen~\cite{DBLP:conf/apn/ValkJ84}.
This will enable us to
prove Lemma~\ref{lem:coreforlinear} in
Section~\ref{sec:ProofLinear}.

\subsection{Computing $\min(\vec{X})$ for $\vec{X}\subseteq\setN^d$}\label{sec:MinElem}\label{sec:VJ}

For $\vec{X}\subseteq \setN^d$ we call
a vector $\vec{m}\in\vec{X}$ \emph{minimal in} $\vec{X}$ if there is
no vector $\vec{x}\in\vec{X}$ such that $\vec{x}\leq \vec{m}$ and
$\vec{x}\neq\vec{m}$.
(We recall that $\vec{x}\leq\vec{y}$ denotes that $\vec{x}(i)\leq \vec{y}(i)$
for all $i\in[1,d]$.)
By $\min(\vec{X})$ we denote \emph{the set of minimal elements in}
$\vec{X}$.
Since $\leq$ is a~well-partial-order on $\setN^d$ (by Dickson's lemma),
the set  $\min(\vec{X})$ is
finite and
for every $\vec{x}\in\vec{X}$ there exists (at least one) $\vec{m}\in\min(\vec{X})$ such that $\vec{m}\leq \vec{x}$.

As a basis for computing  $\min(\vec{X})$ (for special sets
$\vec{X}\subseteq\setN^d$), it is useful to extend the ordered set
$(\setN,\leq)$ with an extra element $\omega\not\in\setN$ so that
$x\leq \omega$ for every $x\in\setN_\omega$, where $\setN_\omega$ denotes $\setN\cup\{\omega\}$.
By $\setN_\omega^d$ we denote the set of $d$-dimensional vectors over
$\setN_\omega$; 
the (component-wise) order $\leq$ on $\setN^d$ is naturally extended
to $\setN_\omega^d$.
For $\vec{v}\in\setN_\omega^d$ we put
$\dreg{\vec{v}}=\{\vec{y}\in\setN^d \mid \vec{y}\leq\vec{v}\}.$
Hence even when $\vec{v}$ has some
$\omega$-components, 
 $\vec{y}\mathop{\in}\dreg{\vec{v}}$ 
has none.

For $\vec{X}\subseteq\setN^d$ we trivially have
$\min(\vec{X})=\min(\vec{X}\mathop{\cap}\dreg{(\omega,\omega,\ldots,\omega)})$.
If we want to describe 
$\min(\vec{X}\mathop{\cap}\dreg{\vec{v}})$, for $\vec{v}\in\setN^d_\omega$,
and we have some  $\vec{y}\in(\vec{X}\mathop{\cap}\dreg{\vec{v}})$,
then we observe that 
\begin{center}
$\min(\vec{X}\mathop{\cap}\dreg{\vec{v}})=
	\min\Big(\{\vec{y}\}\cup
	\min\big(\vec{X}\mathop{\cap}(\dreg{\vec{v}}\smallsetminus\{\vec{x}\mid
	\vec{y}\leq\vec{x}\})\big)\Big).$
\end{center}
To write this more concretely,
by $\vec{v}[i\leftarrow k]$, where $i\in[1,d]$ and $k\in\setN$,
we denote the vector $\vec{v}'\in\setN_\omega^d$ coinciding with $\vec{v}$
except that we have $\vec{v}'(i)=k$, and we put
\begin{center}
	$\delta_{\vec{y}}(\vec{v})=\{\vec{w}\in\setN_\omega^d\mid \vec{w}=\vec{v}[i\leftarrow 
	(\vec{y}(i){-}1)], i\in[1,d], \vec{y}(i)>0\}$. 
\end{center}

\begin{obs}
  For all $\vec{v}\in\setN_\omega^d$ and
  $\vec{y}\in\dreg{\vec{v}}$ we have:
  \begin{enumerate}
  \item			
    Each $\vec{w}\in\delta_{\vec{y}}(\vec{v})$ is strictly less than
    $\vec{v}$ (i.e., $\vec{w}\leq \vec{v}$ and
    $\vec{w}\neq\vec{v}$).
  \item
    $\dreg{\vec{v}}\mathop{\smallsetminus}\{\vec{x}\mid\vec{y}\leq
    \vec{x}\}=\bigcup_{\vec{w}\in\delta_{\vec{y}}(\vec{v})}\dreg{\vec{w}}$.
  \end{enumerate}
\end{obs}

\begin{obs}\label{obs:valkprelim}
  For all $\vec{X}\subseteq \setN^d$, 
  $\vec{v}\in\setN_\omega^d$, and
  $\vec{y}\in(\vec{X}\mathop{\cap}\dreg{\vec{v}}$) we have:
  \begin{center}
    $\min(\vec{X}\mathop{\cap}\dreg{\vec{v}})=
    \min\left(\{\vec{y}\}\cup\bigcup_{\vec{w}\in\delta_{\vec{y}}(\vec{v})}\min(\vec{X}\mathop{\cap}\dreg{\vec{w}})\right).$
  \end{center}		
\end{obs}
\noindent %FIXED indent
Since each strictly decreasing sequence $\vec{v}_0,\vec{v}_1,\vec{v}_2,\ldots$ of
vectors in $\setN^d_\omega$ is finite, we easily observe that there is an algorithm stated
in the next lemma. Its inputs are special algorithms that
we call \emph{set-related algorithms}. Each set-related algorithm is
related to some set
$\vec{X}\subseteq\setN^d$ (for some $d\in\setN$); given
 $\vec{v}\in\setN^d_\omega$, the algorithm decides 
if $(\vec{X}\mathop{\cap}\dreg{\vec{v}})$ is nonempty, and in the
positive case returns some
$\vec{y}\in(\vec{X}\mathop{\cap}\dreg{\vec{v}})$.

\begin{lem}\label{lem:valkjantzen}
  There is an algorithm that, 
  given a~set-related algorithm related to $\vec{X}\subseteq\setN^d$,
  computes the set $\min(\vec{X})$.
\end{lem}	
\begin{rem}
In fact, the algorithm claimed by the lemma does not require to get a code of
	a set-related algorithm; it suffices to get (black-box)
	access to such an algorithm.
\end{rem}	
\vspace*{-\baselineskip} %FIXED Added negative Space to move proof to page

\subsection{Proof of Lemma~\ref{lem:coreforlinear}}\label{sec:ProofLinear}

Now we prove the lemma whose statement is repeated here:
\begin{quote}
  \emph{Given a Petri net $A$ of dimension $d$, and (a presentation
  of) a linear set
  $\vec{L}\subseteq \setN^d$, there is an effectively constructible 
  semilinear non-reachability core $\vec{C}$ for $\vec{L}$.}
\end{quote}
We consider a fixed Petri net $A$ of dimension $d$, and
 we first prove the claim for the case where
$\vec{L}$ is a singleton;
hence $\vec{L}=\{\vec{b}\}$ (there is a
basis $\vec{b}\in\setN^d$, but no periods). 
We observe that if
$\|\vec{x}\|>\|\vec{b}\|$ (where $\|\vec{x}\|=\sum_{i=1}^d
\vec{x}(i)$), 
 then a necessary condition for reachability
of $\vec{b}$ from $\vec{x}$ 
is that $\vec{x}$ belongs to the set 
\begin{center}
  $\DC=\{\vec{x}\in\setN^d\mid$ there is
  $\vec{x}'$ such that $\vec{x}\reachA\vec{x}'$ and
  $\|\vec{x}\|>\|\vec{x}'\|\}$.
\end{center}
For $\vec{x}\in \DC$ we say that $\vec{x}$ \emph{can Decrease the
token-Count}. Since there is no infinite sequence
$\vec{x}_1,\vec{x}_2,\vec{x}_3,\dots$ in $\setN^d$ where 
$\|\vec{x}_1\|>\|\vec{x}_2\|>\|\vec{x}_3\|>\cdots$, 
for 
$\NDC=\overline{\DC}$ (the complement of $\DC$, i.e.
$\setN^d\smallsetminus\DC$)
we note the following trivial fact:

\begin{obs}\label{obs:NDChome}
  $\NDC$ is a home-space for every $\vec{X}\subseteq \setN^d$.
\end{obs}

\noindent %FIXED indent
Proposition~\ref{prop:MINDCconstruct} is a crucial ingredient for 
Proposition~\ref{prop:coreforsingleton} that finishes the proof of Lemma~\ref{lem:coreforlinear}
in the special case when $\vec{L}$ is a singleton.

\begin{prop}\label{prop:MINDCconstruct}
The set $\DC$ is  upward closed and
the set $\min(\DC)$ 
is effectively constructible. Hence both $\DC$ and $\NDC$ are effectively
constructible semilinear sets. 
\end{prop}	
\begin{proof}
  If  $\vec{x}\xrightarrow{\sigma}\vec{x}'$, 
  then
  $\vec{x}+\vec{y}\xrightarrow{\sigma}\vec{x}'+\vec{y}$ (by the 
  monotonicity property of Petri nets). Since 
  $\|\vec{x}\|>\|\vec{x}'\|$ entails
  $\|\vec{x}+\vec{y}\|>\|\vec{x}'+\vec{y}\|$, it is clear that
  $\DC$ is upward closed (i.e., if $\vec{x}\in\DC$ and
  $\vec{x}\leq\vec{y}$, then $\vec{y}\in\DC$).
  
  Regarding the effective constructability of $\min(\DC)$, we
  recall Lemma~\ref{lem:valkjantzen}. The question whether
  $(\DC\mathop{\cap}\dreg{\vec{v}})$ is nonempty, for a given
  $\vec{v}\in\setN_\omega^d$, can be reduced to the reachability
  problem in a standard way (recall the technique sketched for
  Proposition~\ref{prop:decsemilinseq}): We construct a net of bigger dimension from the original net, that first generates some $\vec{y}\in\setN^d$ belonging to $\dreg{\vec{v}}$ that is frozen, and then some $\vec{y}'$ reachable
  from $\vec{y}$ in the original net that is also frozen, and in the final phase a particular place can reach
  zero if, and only if, $\|\vec{y}\|>\|\vec{y}'\|$. 
  Hence in the positive case 
  a~witness of the respective reachability also yields 
  some $\vec{y}\in (\DC\mathop{\cap}\dreg{\vec{v}})$.

  The effective semilinearity of  $\DC$ and $\NDC$ follows trivially.
\end{proof}

\begin{prop}\label{prop:coreforsingleton}
  Given a Petri net $A$ of dimension $d$ and a vector 
  $\vec{b}\in\setN^d$,
  the set
  \begin{center}
$\vec{C}=\NDC\,\cap\big(\{\vec{x}\in\setN^d \mid   \|\vec{x}\|>\|\vec{b}\|\}\cup
    \{\vec{x}\in\setN^d\mid \|\vec{x}\|\leq \|\vec{b}\|$ and $\vec{x}\not\reachA
    \vec{b}\}\big)$
  \end{center}
  is an effectively constructible semilinear non-reachability core for
  $\{\vec{b}\}$.
\end{prop}	
\begin{proof}
  We first show that $\vec{C}$ is a core for
	$\{\vec{b}\}$, i.e.,
	$\vec{C}\subseteq\overline{\pre(\{\vec{b}\})}\subseteq\pre(\vec{C})$:
  \begin{enumerate}
  \item		
    We have	
    $\vec{C}\not\reachA\{\vec{b}\}$,
    since
    $\vec{b}$ is clearly not reachable from any element of
    $\vec{C}$.
  \item		
    For each $\vec{x}\in\setN^d$, if $\vec{x}\not\reachA\vec{b}$, then 
    $\vec{x}\reachA \vec{x}'\not\reachA \vec{b}$ for some
    $\vec{x}'\in\NDC$ (recall
    Observation~\ref{obs:NDChome}); 
    the facts $\vec{x}'\in\NDC$ and $\vec{x}'\not\reachA \vec{b}$
    obviously entail $\vec{x}'\in\vec{C}$, and thus
    $\vec{x}\reachA\vec{C}$.
  \end{enumerate}
  \noindent %FIXED indent
  The effective semilinearity of $\vec{C}$ follows from
  Proposition~\ref{prop:MINDCconstruct} and from the fact
  that the finite set 
  $\{\vec{x}\in\setN^d\mid \|\vec{x}\|\leq \|\vec{b}\|$ and $\vec{x}\not\reachA
  \vec{b}\}$ can be constructed by repeatedly using an algorithm deciding
  reachability.
\end{proof}

\noindent %FIXED indent
Now we proceed to prove Lemma~\ref{lem:coreforlinear} in general.
We have a Petri net $A$ of dimension $d$, and a linear set
$\vec{L}$ presented by a basis $\vec{b}\in\setN^d$ and periods
$\vec{p}_1,\vec{p}_2\dots,\vec{p}_k\in\setN^d$;
we aim to
construct a semilinear non-reachability core for $\vec{L}$.
We would like
to generalize the above special-case proof (which is, in fact, closely related 
to the approach in~\cite{EscrigJohnen89}),
with the upward closed set $\DC$. But here is a subtle
problem that leads us to
not working with
configurations $\vec{x}\in\setN^d$ directly but rather via
their $\vec{L}$-like presentations.

We note that each configuration $\vec{x}\in\setN^d$ can be presented
as 
\begin{center}
  $\vec{x}=\vec{y}+\vec{u}(1)\cdot\vec{p}_1
  +\vec{u}(2)\cdot\vec{p}_2\cdots
  +\vec{u}(k)\cdot\vec{p}_k$ 
\end{center}
for at least one (but often more)
pairs $(\vec{y},\vec{u})\in\setN^d\times\setN^k$.
For $\vec{y}\in\setN^d$ and $\vec{u}\in\setN^k$ we put
\begin{center}
  $\prmark(\vec{y},\vec{u})=\vec{y}+\vec{u}(1)\cdot\vec{p}_1
  +\vec{u}(2)\cdot\vec{p}_2\cdots
  +\vec{u}(k)\cdot\vec{p}_k$.
\end{center}
Hence 
$\vec{L}=\{\prmark(\vec{b},\vec{u}) \mid \vec{u}\in\setN^k\}$.

Let $\DCB$ (determined by the Petri net $A$ and the
sequence of periods of $\vec{L}$) be the set of
presentation pairs that present 
configurations that \emph{can Decrease the token-Count in the
presentation Basis}:
\begin{center}
  $\DCB=\{(\vec{y},\vec{u})\in\setN^d\times\setN^k\mid \exists
  (\vec{y}',\vec{u}'): 	\|\vec{y}\|>\|\vec{y}'\|,
  \prmark(\vec{y},\vec{u})\reachA
  \prmark(\vec{y}',\vec{u}')\}$.
\end{center}
We note that if $\vec{y}\geq \vec{p}_i$, for
some $i\in[1,k]$, then we trivially have $(\vec{y},\vec{u})\in
\DCB$ since 
$\prmark(\vec{y},\vec{u})=\prmark(\vec{y}-\vec{p}_i,\vec{u}')$
where $\vec{u}'$ arises from $\vec{u}$ by
adding $1$ to $\vec{u}(i)$. (As expected, we assume that all
$\vec{p}_i$ are nonzero vectors.)

\begin{prop}\label{prop:DCBeffective}
  $\DCB$ is upward closed 
  and the set $\min(\DCB)$ is
  effectively constructible.
\end{prop}
\begin{proof}
  As expected, we compare the elements of $\DCB$ 
  component-wise. 
  To show that $\DCB$ is upward closed, we assume that
  $(\vec{y}_1,\vec{u}_1)\in\DCB$ and
  $(\vec{y}_1,\vec{u}_1)\leq (\vec{y}_2,\vec{u}_2)$. 
  To demonstrate that $(\vec{y}_2,\vec{u}_2)\in\DCB$ as well, we again use
  monotonicity of Petri nets:
  Since $\prmark(\vec{y}_1,\vec{u}_1)\xrightarrow{\sigma}\prmark(\vec{y}'_1,\vec{u}'_1)$
  (for some sequence $\sigma$)
  where $\|\vec{y}_1\|>\|\vec{y}'_1\|$, and
  $\prmark(\vec{y}_1,\vec{u}_1)\leq\prmark(\vec{y}_2,\vec{u}_2)$,
  we have
  $\prmark(\vec{y}_2,\vec{u}_2)\xrightarrow{\sigma}\prmark(\vec{y}'_1{+}(\vec{y}_2{-}\vec{y}_1),\vec{u}'_1{+}(\vec{u}_2{-}\vec{u}_1))$;	
  $\|\vec{y}_1\|>\|\vec{y}'_1\|$ entails 	
  $\|\vec{y}_2\|>\|\vec{y}'_1{+}(\vec{y}_2{-}\vec{y}_1)\|$.
  
  The effective constructability of $\min(\DCB)$ is again based
  on Lemma~\ref{lem:valkjantzen}, when we identify
  $\setN^d\times\setN^k$ with $\setN^{d+k}$. It is again a technical routine
  to show that the question whether
  $(\DCB\,\mathop{\cap}\dreg{\vec{v}})$ is nonempty, for a given
  $\vec{v}\in\setN_\omega^{d+k}$, can be reduced to the reachability
  problem, so that 
  in the positive case a witness of this reachability also yields
  some $(\vec{y},\vec{u})\in
  (\DCB\,\mathop{\cap}\dreg{\vec{v}})$. 
\end{proof}
\noindent %FIXED indent
We now define the set of configurations with presentations in which
the basis cannot be decreased:

\begin{center}
  $\NDCBConf=\{\vec{x}\in\setN^d\mid
  \vec{x}=\prmark(\vec{y},\vec{u})$ for some
  $(\vec{y},\vec{u})\not\in \DCB\}$.
\end{center}
\begin{obs}\label{obs:NDCBhome}
  $\NDCBConf$ is a home-space for every $\vec{X}\subseteq \setN^d$.
\end{obs}
\begin{proof}
  Suppose there is some $\vec{x}\in\setN^d$ such that
$\vec{x}\not\reachA\NDCBConf$; we fix one such $\vec{x}$ that can be
written as $\vec{x}=\prmark(\vec{y},\vec{u})$ for $\vec{y}$ with the
least norm $\|\vec{y}\|$.
Since $\vec{x}\not\in\NDCBConf$, we have
$(\vec{y},\vec{u})\in\DCB$, which entails a contradiction by the
definition of $\DCB$.
\end{proof}

\begin{prop}\label{prop:NCDBsemilin}
  $\NDCBConf$ is an effectively constructible semilinear set.
\end{prop}
\begin{proof}
  By Proposition~\ref{prop:DCBeffective}, $\DCB$ is an effectively constructible semilinear
  set. Since semilinear sets (effectively) coincide with the sets definable in
  Presburger arithmetic, the claim is clear.
\end{proof}
\noindent %FIXED indent
The next proposition finishes a proof of Lemma~\ref{lem:coreforlinear}, and thus also 
of Theorem~\ref{th:semhomspace}.

\begin{prop}\label{prop:coreforlinear}
  Given a Petri net $A$ of dimension $d$ and a linear set
  $\vec{L}\subseteq\setN^d$ presented by
  $(\vec{b},\vec{p}_1,\vec{p}_2,\ldots,\vec{p}_k)$,
  the set
  \begin{center}
    $\vec{C}=\{\vec{x}\in\setN^d\mid \vec{x}=\prmark(\vec{y},\vec{u})$
    where $(\vec{y},\vec{u})\not\in\DCB$ and
    \\
    either $\|\vec{y}\|>\|\vec{b}\|$, or $\|\vec{y}\|\leq \|\vec{b}\|$ and 
    $\prmark(\vec{y},\vec{u})\not\reachA \vec{L}\}$
  \end{center}
  is an effectively constructible semilinear non-reachability core for
  $\vec{L}$.
\end{prop}	
\begin{proof}
  We note that $\vec{C}$ is a subset of $\NDCBConf$, and we recall
  that 
	$\vec{x}\in \vec{L}$ if{f} $\vec{x}=\prmark(\vec{b},\vec{u})$ for some
  $\vec{u}\in\setN^k$. 
  We verify that $\vec{C}$ 
  is a core for 
	$\vec{L}$, i.e.,
	$\vec{C}\subseteq\overline{\pre(\vec{L})}\subseteq\pre(\vec{C})$:
  \begin{enumerate}
  \item 
    By definition of $\vec{C}$ we clearly have
    $\vec{C}\not\reachA\vec{L}$.
  \item		
    For each $\vec{x}\in\setN^d$, if $\vec{x}\not\reachA\vec{L}$, then 
    $\vec{x}\reachA \vec{x}'\not\reachA \vec{L}$ for some
    $\vec{x}'\in\NDCBConf$ (recall
    Observation~\ref{obs:NDCBhome}); 
    the facts $\vec{x}'\in\NDCBConf$ and $\vec{x}'\not\reachA \vec{L}$
    obviously entail $\vec{x}'\in\vec{C}$, and thus
    $\vec{x}\reachA \vec{C}$.
  \end{enumerate}
  \noindent %FIXED indent
  Now we aim to show that $\vec{C}$ is an effectively constructible
  semilinear set. 
  We recall Propositions~\ref{prop:NCDBsemilin}
  and~\ref{prop:DCBeffective},
  and the fact that
  for any concrete $\vec{y}$ and $\vec{u}$ we can decide if
  $\prmark(\vec{y},\vec{u})\reachA \vec{L}$.
  Though 
  there are only finitely many $\vec{y}$ to consider, namely those
  satisfying $\|\vec{y}\|\leq\|\vec{b}\|$, we are not
  done:
  it is not immediately obvious how to express
  $\prmark(\vec{y},\vec{u})\not\reachA \vec{L}$ in Presburger arithmetic,
  even when $\vec{y}$ is fixed. To this aim, for any fixed
  $\vec{y}\in\setN^d$ we
  define the set
  \begin{center}
    $\vec{U}_{\vec{y}}=\{\vec{u}\in\setN^k \mid \prmark(\vec{y},\vec{u})\reachA
    \vec{L}\}=
    \{\vec{u}\in\setN^k \mid \exists \vec{u}'\in\setN^k: 
    \prmark(\vec{y},\vec{u})\reachA \prmark(\vec{b},\vec{u}')\}$.
  \end{center}
  For each fixed $\vec{y}\in\setN^d$,
  the set $\vec{U}_{\vec{y}}$ is clearly upward closed 
  (by monotonicity of Petri nets). 
  Moreover, the set
  $\min(\vec{U}_{\vec{y}})$ is
  effectively constructible, again by using
  Lemma~\ref{lem:valkjantzen}:
  Given a~fixed $\vec{y}$, for each $\vec{v}\in\setN^k_\omega$ we can
  decide whether
  $(\vec{U}_{\vec{y}}\mathop{\cap}\dreg{\vec{v}})$ is nonempty
  by a reduction to the reachability problem, so that 
  in the positive case a witness of this reachability 
  also yields some
  $\vec{u}\in(\vec{U}_{\vec{y}}\mathop{\cap}\dreg{\vec{v}})$.
  
  Now it is clear that we can effectively construct a Presburger
  formula defining $\vec{C}$; hence $\vec{C}$ is a semilinear set 
  for which we can effectively construct a presentation.
\end{proof}

\section{Minimal Reachable Configurations}\label{sec:Ackone}

	We have proven the decidability
	(Theorem~\ref{th:semhomspace}), and now we aim to analyze the
	presented approach to get some complexity upper bounds that will
	enable us to prove Theorem~\ref{th:AckerComplex}.
This aim leads us to show several Ackermannian-time algorithms
in this section.

The first algorithm gets
 a~Petri net $A$ of dimension $d$ and a configuration $\vec{x}\in\setN^d$
 as input,
and computes the set $\min(\post(\vec{x}))$, i.e. the set of 
minimal configurations in the respective reachability set.
The second algorithm computes
$\min(\post(\vec{x})\cap\vec{S})$ when it gets 
(a presentation of) a semilinear set  $\vec{S}\subseteq\setN^d$
besides $A$ and $\vec{x}$.
The third algorithm gets $A,\vec{x}$, and 
(a presentation of) a semilinear predicate $P\subseteq\setN^h\times\setN^d\times \setN^d$
(for some $h\in\setN$), and 
computes the set
\[\min(\{\vec{x}\in\setN^h \mid \exists \alpha,\beta\in\setN^d:
\alpha\xrightarrow{A^*}\beta\wedge (\vec{x},\alpha,\beta)\in P\}).\]
\noindent %FIXED indent
The complexity of computing the above mentioned minimal configurations can be derived by
using the approach by Hsu-Chun Yen and Chien-Liang Chen
in~\cite{DBLP:journals/tcs/YenC09}; they observed that complexity bounds
on a set-related algorithm related to some set $\vec{X}\subseteq \setN^d$ (recall the
definition before Lemma~\ref{lem:valkjantzen})
allow us to derive complexity bounds on the computation of
$\min(\vec{X})$. 
As a crucial ingredient here, 
we recall the known complexity
upper bound for reachability in Section~\ref{sec:fixeddim}.
In Section~\ref{sec:minreachconfig} we derive an~Ackermannian bound on the
size of minimal configurations in Petri net reachability sets, and we
extend this bound in Section~\ref{sec:exttosemilin} and
in Section~\ref{sec:exttosemilinpred} to obtain the mentioned second
algorithm
and the third algorithm, respectively.

\begin{rem}
Mayr and Meyer described in~\cite{DBLP:journals/jacm/MayrM81} a family
of Petri nets that exhibits finite reachability sets whose size 
grows as the Ackermann function; hence also the size
of the maximal configurations in these sets grows similarly.
Concerning the size of minimal configurations, we cannot deduce any
interesting size properties using the same family. However, by using
the family of Petri nets recently introduced
in~\cite{DBLP:conf/focs/Leroux21,DBLP:conf/focs/CzerwinskiO21,DBLP:conf/stacs/000122}
for proving that the reachability problem is Ackermann-hard, we can
observe that the maximal size of minimal configurations in Petri net reachability sets
grows at least as the Ackermann function.
\end{rem}

\subsection{Petri Net Reachability Problem in Fixed Dimension}\label{sec:fixeddim}

Here we recall some definitions in order to state that the Petri net
reachability problem is primitive-recursive when
restricted to a~fixed dimension, and 
Ackermannian in general.

The \emph{fast-growing functions} $F_d \colon \setN \to \setN$,
$d\in\setN$, are
defined inductively as follows:
\begin{center}
$F_0(n)=n+1$,
and $F_{d+1}(n)=F_d^{(n+1)}(n)$;
\end{center}
	where by $f^{(n)}$, for a function
$f\colon \setN\to\setN$, we mean the respective iteration of $f$ (i.e., $f^{(n+1)}=f^{(n)}\circ f$). Following \cite{Schmitz16toct}, we introduce the class $\bold{F}_d$ of
functions computable in time $O(F_d(F_{d-1}^{(c)}(n)))$ where $n$ is the size of the input and $c\in\setN$ is any constant. 
We recall that $\bigcup_{d\in\setN}\bold{F}_d$ is the class of 
\emph{primitive-recursive functions}. 
We also introduce the function  $F_\omega\colon \setN\to\setN$
defined by $F_\omega(n)=F_n(n)$, which is a variant of the Ackermann
function;
by $\bold{F}_\omega$ we denote the class of
functions computable
in time $O(F_\omega(F_{d}(n)))$ where $d\in\setN$ is any constant
and $n$ is the size of the input. 
A function in $\bold{F}_\omega$ is said to be computable in \emph{Ackermannian time}.
(We note that Ackermannian time coincides with Ackermannian space.)

For $\vec{x}\in\setN^d$ 
we have defined the norm of $\vec{x}$ as 
$\|\vec{x}\|=\sum_{i=1}^d\vec{x}(i)$.
Now we extend the notion of norm to other objects.
For a \emph{Petri net action} $a=(\vec{a}_-,\vec{a}_+)$, by its \emph{norm}
we mean $\norm{a}=\max\{\norm{\vec{a}_-},\norm{\vec{a}_+}\}$. 
For a~\emph{Petri net} $A$, by its \emph{norm} we mean
$\norm{A}=\max_{a\in A}\norm{a}$. The \emph{norm} of a linear set
$\vec{L}\subseteq\setN^d$ implicitly given by a presentation
$(\vec{b},\vec{p}_1,\vec{p}_2,\ldots,\vec{p}_k)$ is defined by
$\norm{\vec{L}}=\max\{\norm{\vec{b}},\norm{\vec{p}_1},\norm{\vec{p}_2},\ldots,\norm{\vec{p}_k}\}$.
The \emph{norm} of a semilinear set $\vec{S}\subseteq\setN^d$
implicitly given by a sequence of presentations of $\vec{L}_1,\vec{L}_2,\ldots,\vec{L}_m$ is defined by $\norm{\vec{S}}=\max_{1\leq n\leq m}\norm{\vec{L}_n}$.

Now we recall a result showing 
that the reachability problem restricted to Petri nets of dimension
$d$ is in $\bold{F}_{d+4}$, and that the general Petri net
reachability problem is in $\bold{F}_\omega$. (We view a decision problem
as a function with the co-domain $\{0,1\}$.) This result is crucial
for us to derive the upper bound in Theorem~\ref{th:AckerComplex}.
\begin{thmC}[\cite{DBLP:conf/lics/LerouxS19}]\label{thm:reach} %FIXED Removed Double Paranthesis
  There is a constant $c>0$ such that 
  for al{l} $d,n,A,\vec{x},\vec{y}$ 
  where  $d,n\in\setN$, $A$ is a Petri net of dimension $d$,
  $\vec{x},\vec{y}\in\setN^d$, and the norms of
  $A,\vec{x},\vec{y}$ are bounded by $n$, we have that
  if $\vec{x}\xrightarrow{A^*}\vec{y}$, then $\vec{x}\xrightarrow{\sigma}\vec{y}$
  for a word $\sigma\in A^*$ such that $|\sigma|\leq F_{d+4}\circ
  F_{d+3}^{(c)}(n)$.
\end{thmC}

\begin{center}%FIXED Aligned The Beginning quote in the middle
We remark that in what follows we formulate some results in the form \par
  ``There is a constant $c'>0$
  such that...''
\end{center}
Naturally we could replace $c'$ with $c$ without
changing the meaning of the respective statements,
but we prefer keeping the difference in order to highlight 
the special role of the constant $c$ introduced in Theorem~\ref{thm:reach}.

\subsection{Minimal Reachable Configurations}\label{sec:minreachconfig}
We provide an algorithm computing the set of minimal reachable configurations,
by following the approach of~\cite{DBLP:journals/tcs/YenC09}.
To ease notation, we introduce the functions $f_d=F_{d+4}\circ
F_{d+3}^{(c)}$ ($d\in\setN$) where $c$ is the constant introduced in
Theorem~\ref{thm:reach}, and we first prove the following proposition;
for $\vec{v}\in\setN_\omega^d$, by its \emph{norm}
we mean $\norm{\vec{v}}=\sum_{i: \vec{v}(i)\not=\omega}\vec{v}(i)$.
\begin{prop}\label{lem:findone}
  For al{l} $d$, $n$, $A$, $\vec{x}$, $\vec{v}$, where $d,n\in\setN$,
  $A$ is a Petri net of dimension $d$,
  $\vec{x}\in\setN^d$, $\vec{v}\in\setN^d_\omega$, and
  the norms of $A,\vec{x},\vec{v}$ are bounded by $n$,
  we have that
	if $(\post(\vec{x})\mathop{\cap}\reg{\vec{v}})$ is nonempty, then there
	is $\vec{y}\in (\post(\vec{x})\mathop{\cap}\reg{\vec{v}})$
	such that $\vec{x}\xrightarrow{\sigma}\vec{y}$ for some
	$\sigma\in A^*$ where $|\sigma|\leq f_d(n)$.
\end{prop}
\begin{proof}
  For $n=0$
  the claim is trivial, so we assume $n\geq 1$.
  
  For each $j\in[1,d]$ we define 
  the Petri net action $b_j=(\vec{i}_j,\vec{0})$ where
  $\vec{i}_j(j)=1$ and 	$\vec{i}_j(i)=0$ for all
  $i\in[1,d]\smallsetminus\{j\}$;
  this action decrements the $j$th component of configurations. 
  We put $I_\omega=\{j\mid j\in[1,d], \vec{v}(j)=\omega\}$, and by $B$ we denote
  the Petri net $\{b_j\mid j\in I_\omega\}$. Since $n\geq 1$, we derive $\norm{A\cup B}\leq n$.

  Let us now consider a configuration
  $\vec{z}\in(\post(\vec{x})\mathop{\cap}\reg{\vec{v}})$.
  Let $\vec{c}$ be the configuration arising from $\vec{z}$ by replacing
  the components in $I_\omega$ with zero; we thus have  $\norm{\vec{c}}\leq \norm{\vec{v}}\leq n$
  (using the fact that $\vec{c}\leq\vec{z}$, and thus $\vec{c}\,\mathop{\in}\reg{\vec{v}}$).
  
  From $\vec{x}\xrightarrow{A^*}\vec{z}$ and $\vec{z}\xrightarrow{B^*}\vec{c}$ we derive $\vec{x}\xrightarrow{(A\cup B)^*}\vec{c}$. 
  By Theorem~\ref{thm:reach}
  we deduce that $\vec{x}\xrightarrow{u}\vec{c}$ for some word
  $u\in (A\cup B)^*$ for which $|u|\leq f_d(n)$. Since Petri net
  actions in $B$ only decrease some components, 
  we can assume that all these actions in $u$ are at the end; hence 
  $u=\sigma v$ where $\sigma\in A^*$ and $v\in B^*$, and we have
  $\vec{x}\xrightarrow{\sigma}\vec{y}\xrightarrow{v}\vec{c}$ for a
  configuration $\vec{y}\in\post(\vec{x})$.
  Since $\vec{c}\leq\vec{z}$, $\vec{z}\in\reg{\vec{v}}$, and 
  $\vec{y}\xrightarrow{v}\vec{c}$ only decreases the components that are
  $\omega$ in $\vec{v}$, we deduce that $\vec{y}\in \reg{\vec{v}}$.
\end{proof}
\noindent %FIXED indent
To ease the formulation of the next proposition, for all
$d\in\setN$
we define the functions $g_d\colon \setN\to\setN$ by
\begin{center}
$g_d(n)=n\cdot(\,2+f_d(n)\,)$.
\end{center}
\begin{prop}\label{lem:mainreach}
  For al{l} $d,n,A,\vec{x},\vec{v},\vec{m}$, where $d,n\in\setN$,
  $A$ is a Petri net of dimension $d$,  
  $\vec{x}\in\setN^d$, $\vec{v}\in\setN^d_\omega$,
  $\vec{m}$ belongs to $\min(\post(\vec{x})\mathop{\cap}\reg{\vec{v}})$,
  and the norms of $A,\vec{x},\vec{v}$ are bounded by $n$,
  there exists a word $\sigma\in A^*$ 
  such that  $\vec{x}\xrightarrow{\sigma}\vec{m}$ and
  $|\sigma|\leq f_d\circ
  g_d^{(k)}(n)$ where
  $k=|\{i\mid \vec{v}(i)=\omega\}|$.
\end{prop}
\begin{proof}
  The strict version $<$ of the relation $\leq$ on $\setN_\omega^d$ (defined by $\vec{w}<\vec{v}$ if $\vec{w}\leq \vec{v}$ and $\vec{w}\not=\vec{v}$) is clearly well-founded. We use this property for an inductive proof.

  We aim to show the claim for a considered tuple
  $d,n,A,\vec{x},\vec{v},\vec{m}$, while we can assume that the
  claim is valid for $d,n',A,\vec{x},\vec{w},\vec{m}'$ for all $\vec{w}<\vec{v}$ and all
  $\vec{m}'\in\min(\post(\vec{x})\mathop{\cap}\reg{\vec{w}})$.
  
  Since $\vec{m}$ is in
  $(\post(\vec{x})\mathop{\cap}\reg{\vec{v}})$, we deduce
  from Lemma~\ref{lem:findone} that we can fix 
  $\vec{y}\in(\post(\vec{x})\mathop{\cap}\reg{\vec{v}})$ 
  and a word $\sigma\in A^*$ such that
  $\vec{x}\xrightarrow{\sigma}\vec{y}$ and
  $|\sigma|\leq f_d(n)$; we thus have 
  $\norm{\vec{y}}\leq \norm{\vec{x}}+\norm{\vec{A}}\cdot|\sigma|\leq g_d(n)-n$.
  If $\vec{m}=\vec{y}$, then the claim is proved; so we assume that $\vec{m}\not=\vec{y}$. 
  
  By Observation~\ref{obs:valkprelim}
  we can fix $\vec{w}\in\delta_{\vec{y}}(\vec{v})$ such that
  $\vec{m}\in\min(\post(\vec{x})\mathop{\cap}\reg{\vec{w}})$;
  since $\vec{w}\in\delta_{\vec{y}}(\vec{v})$, we have
  $\vec{w}<\vec{v}$.
  By the induction hypothesis, there is a word $\sigma'\in A^*$
  such that $\vec{x}\xrightarrow{\sigma'}\vec{m}$ and
  $|\sigma'|\leq f_d\circ g_d^{(k')}(n')$ where $n'=\max\{\norm{A},\norm{\vec{x}},\norm{\vec{w}}\})$ and $k'=|\{i\mid \vec{w}(i)=\omega\}|$.
  Putting $k=|\{i\mid \vec{v}(i)=\omega\}|$, we observe that $k'=k$ or $k'=k-1$. If $k'=k$,
  then $\norm{\vec{w}}<\norm{\vec{v}}$ and we are done by 
  monotonicity of $f_d$ and $g_d$. 
  Otherwise $k'=k-1$ and in
  that case $\norm{\vec{w}}\leq \norm{\vec{v}}+\norm{\vec{y}}\leq g_d(n)$ since in that case $\vec{w}$ is obtained from $\vec{v}$ by replacing component $i$ of $\vec{v}$ for some $i$ such that $\vec{v}(i)=\omega$ and $\vec{y}(i)>0$ by $\vec{y}(i)-1$. 
  It follows that
  $n'\leq g_d(n)$
  and we are done also in that case by monotonicity of $f_d$ and $g_d$.
\end{proof}

\noindent %FIXED indent
Finally, by instantiating the previous proposition with
$\vec{v}=(\omega,\omega,\ldots,\omega)$, and by bounding $f_d\circ
g_d^{(d)}(n)$ as provided by the next proposition, we deduce the following two corollaries.
\begin{prop}\label{prop:Fbound}
  For every $d,n$, we have $f_d\circ g_d^{(d)}(n)\leq F_{d+5}((d+1+n)(c+2))$.
\end{prop}
\begin{proof}
  As $F_2(x)=2^x(x+1)-1$ we deduce that $F_2(x)\geq x(2+x)$ for every $x\geq 0$. It follows from $F_{d+4}(x)\geq F_2(x)$ that $F_{d+4}(x)\geq x(2+x)$ for every $x$. Now, let $y\geq 0$ and let us put $x=F_{d+4}^{(c+1)}(y)$. We have $F_{d+4}^{(c+2)}(y)=F_{d+4}(x)\geq x(2+x)\geq y(2+x)$. Since $x= F_{d+4}\circ F_{d+4}^{(c)}(y)\geq F_{d+4}\circ F_{d+3}^{(c)}(y)=f_d(y)$, we deduce that $x\geq f_d(y)$. Combined with $F_{d+4}^{(c+2)}(y)\geq y(2+x)$ we get $F_{d+4}^{(c+2)}(y)\geq y(2+f_d(y))=g_d(y)$. We have proved that $g_d(y)\leq F_{d+4}^{(c+2)}(y)$ for every $y$. In particular $f_d\circ g_d^{(d)}(n)\leq F_{d+4}^{(c+1)}\circ F_{d+4}^{(d(c+2))}(n)\leq F_{d+4}^{((d+1)(c+2))}(n)$ for every $n$. It follows that $f_d\circ g_d^{(d)}(n)\leq F_{d+4}^{((d+1)(c+2))}(n)\leq F_{d+4}^{((d+1+n)(c+2))}((d+1+n)(c+2))= F_{d+5}((d+1+n) (c+2))$.
\end{proof}

\begin{cor}\label{cor:vidal}
  There is a constant $c>0$ such that
  for al{l} $d,n,A,\vec{x},\vec{m}$, where $d,n\in\setN$,
  $A$ is a Petri net of dimension $d$,  
  $\vec{x}\in\setN^d$,
  $\vec{m}$ belongs to $\min(\post(\vec{x}))$,
  and the norms of $A,\vec{x}$ are bounded by $n$,
  there exists a word $\sigma\in A^*$ 
  such that  $\vec{x}\xrightarrow{\sigma}\vec{m}$ and
  $|\sigma|\leq F_{d+5}((d+1+n)(c+2))$.
\end{cor}

\begin{cor}\label{cor:vidal-compute}
  There is a constant $c>0$ such that
  for al{l} $d,n,A,\vec{x}$, where $d,n\in\setN$,
  $A$ is a~Petri net of dimension $d$,  
  $\vec{x}\in\setN^d$,
  and the norms of $A,\vec{x}$ are bounded by $n$,
  the set $\min(\post(\vec{x}))$ is computable in time exponential in
	$F_{d+5}((d+1+n)(c+2))$ and the norms of vectors in that set are bounded by $n\cdot (1+F_{d+5}((d+1+n)(c+2)))$.
\end{cor}
\begin{proof}
  In fact, the set of minimal reachable configurations can be obtained
  by exploring configurations reachable from $\vec{x}$ by
  sequences of at most $F_{d+5}((d+1+n)(c+2))$ actions in $A$. We note that
  the norms of
  configurations reachable in this way are bounded by $\norm{\vec{x}}+F_{d+5}((d+1+n)(c+2))\cdot\norm{A}\leq n\cdot (1+F_{d+5}((d+1+n)(c+2)))$.
\end{proof}

\subsection{Extension to Semilinear Sets}\label{sec:exttosemilin} 
The algorithm computing minimal reachable configurations can
be also simply used for computing the set $\min(\post(\vec{x})\cap \vec{S})$ where
$\vec{S}$ is a semilinear set; we thus formulate this fact as a
corollary (though with a proof).
We recall that the norm of a semilinear set is the
maximum norm of vectors occurring in its (implicitly assumed) presentation.
\begin{cor}\label{cor:semi}
  There is a constant $c>0$ such that
  for al{l} $d,n,A,\vec{x},\vec{S}$, where $d,n\in\setN$,
  $A$ is a Petri net of dimension $d$,  
  $\vec{x}\in\setN^d$,
  $\vec{S}$ is (a presentation of) a semilinear set $\vec{S}\subseteq \setN^d$,
  and the norms of $A,\vec{x},\vec{S}$ are bounded by $n$,
  the set $\min(\post(\vec{x})\cap\vec{S})$ is computable in time
  exponential in $F_{2d+6}(n(c+2))$ and the norms of vectors in
  that set are bounded by $n\cdot (1+F_{2d+6}((2d+2+n)(c+2)))$.
\end{cor}
\begin{proof}
  Let us consider a $d$-dimensional Petri net $A$, an initial
  configuration $\vec{x}$, and a semilinear set
  $\vec{S}\subseteq \setN^d$ given as the union of linear sets $\vec{L}_1,\vec{L}_2,\ldots,\vec{L}_m$. Since $\min(\post(\vec{x})\cap\vec{S})=\min(\bigcup_{j=1}^m\min(\post(\vec{x})\cap\vec{L}_j))$ we can reduce the problem of computing $\min(\post(\vec{x})\cap\vec{S})$ to the special case of a linear set $\vec{S}$, denoted as $\vec{L}$ in the sequel. So, let $\vec{L}$ be a linear set presented by a basis $\vec{b}\in\setN^d$ and a sequence of periods $\vec{p}_1,\vec{p}_2,\ldots,\vec{p}_k\in\setN^d$, and let us provide an algorithm for computing $\min(\post(\vec{x})\cap\vec{L})$.
  
  To do so, we build from $A$ a new Petri net $B$ of dimension $2d+1$
  defined as follows and an initial configuration
  $(\vec{x},1,\vec{0})$. We associate to each Petri net action
  $a\in A$ of the form $(\vec{a}_-,\vec{a}_+)$ the action
  $((\vec{a}_-,1,\vec{0}),(\vec{a}_+,1,\vec{0}))$ in $B$ that
  intuitively executes $a$ on the first $d$ counters and check
  that the middle counter (the counter $d+1$) is at least $1$.
  We also add in $B$ for each $j\in[1,k]$ an action
  $((\vec{p}_j,0,\vec{0}),(\vec{0},0,\vec{p}_j))$ that removes
  the period $\vec{p}_j$ on the first $d$ counters and adds it
  on the last $d$ counters. Finally, we add to $B$ the action
  $((\vec{b},1,\vec{0}),(\vec{0},0,\vec{b}))$ that decrements
  the middle counter and simultaneously removes $\vec{b}$ from
  the first $d$ counters, and adds $\vec{b}$ on the last $d$ counters. Since for any set $\vec{X}\subseteq \setN^d$ and any set $I\subseteq [1,d]$, the set $\min(\{\vec{x}\in\vec{X}\mid \bigwedge_{i\in I}\vec{x}(i)=0\})$ is equal to $\{\vec{m}\in\min(\vec{X})\mid \bigwedge_{i\in I}\vec{m}(i)=0\})$, one can observe that $\{\vec{0}\}\times\{0\}\times \min(\post(\vec{x})\cap\vec{L})$ is equal to $\min(\textsc{post}^*_B(\vec{x},1,\vec{0}))\cap(\{\vec{0}\}\times\{0\}\times\setN^d)$. 
\end{proof}

\subsection{Extension to Semilinear Predicates}\label{sec:exttosemilinpred} 
By another corollary (with a proof) we also note that the algorithm computing minimal reachable configurations can be used for computing minimal vectors in sets of the following form
\begin{equation}\label{eq:X}
  \vec{X}=\{\vec{x}\in\setN^h \mid \exists \alpha,\beta\in\setN^d: \alpha\xrightarrow{A^*}\beta\wedge (\vec{x},\alpha,\beta)\in P\}
\end{equation}
where $P\subseteq\setN^h\times\setN^d\times\setN^d$ is a semilinear predicate given by a presentation. Notice that we use Greek letters $\alpha$ and $\beta$ in the definition of $\vec{X}$ in order to emphasise vectors that act as configurations of the Petri net $A$.

\begin{cor}\label{cor:pred}
  There is a constant $c>0$ such that
  for al{l} $d,h,n,A,P$, where $d,h,n\in\setN$,
  $A$ is a Petri net of dimension $d$,  
  $\vec{x}\in\setN^d$,
  $P$ is (a presentation of) a semilinear predicate $P\subseteq \setN^h\times\setN^d\times\setN^d$,
  and the norms of $A,\vec{x},P$ are bounded by $n$,
  the set of minimal elements of the set $\vec{X}$ denoted by
  equation~(\ref{eq:X}) is computable in time exponential in
  $F_{2h+4d+6}((2h+4d+2+n)(c+2))$ and the norms of these minimal elements are bounded by $n\cdot (1+F_{2h+4d+6}(n(c+2)))$.
\end{cor}
\begin{proof}
  We first introduce the set $Y$ defined as $Z\cap P$ where 
  \[Z=\{(\vec{x},\alpha,\beta)\in
  \setN^h\times\setN^d\times\setN^d \mid
  \alpha\xrightarrow{A^*}\beta\}.\]
  Since $\min(\vec{X})=\min \{\vec{x}\in\setN^k \mid \exists \alpha,\beta\in\setN^d: ~(\vec{x},\alpha,\beta)\in \min(Y)\}$ it is sufficient to provide an algorithm computing $\min(Y)$.

  Our algorithm is based on the fact that $Z$ is the reachability set
  of a $(h+2d)$-dimensional Petri net $B$ starting from the zero
  configuration and defined as follows from $A$. By $\vec{i}_i$
  we denote the vector in $\setN^h$ defined by $\vec{i}_i(i)=1$
  and $\vec{i}_i(j)=0$ if $j\in[1,h]\backslash\{i\}$. The Petri
  net $B$ is defined as the actions
  $((\vec{0},\vec{0},\vec{0}),(\vec{i}_j,\vec{0},\vec{0}))$
  where $j\in [1,h]$ that increment the counters corresponding
  to $\vec{x}$, actions
  $((\vec{0},\vec{0},\vec{0}),(\vec{0},\vec{i}_j,\vec{i}_j))$
  that increment simultaneously by the same amount the counters
  corresponding to $\alpha$ and $\beta$, and actions obtained
  from $A$ that simulate the computation of $A$ on the counters
  $\beta$ and defined for each action $a$ of $A$ of the form
  $(\vec{a}_-,\vec{a}_+)$ by the action
  $((\vec{0},\vec{0},\vec{a}_-),(\vec{0},\vec{0},\vec{a}_+))$ in
  $B$. Notice that
  $Z=\textsc{post}^*_B(\vec{0},\vec{0},\vec{0})$ and we are done
  by Corollary~\ref{cor:semi}.
\end{proof}

\section{Complexity of the Semilinear Home-Space Problem}\label{sec:Acktwo}
In this section we provide an Ackermannian complexity upper-bound for
deciding the semilinear home-space problem; 
Theorem~\ref{th:AckerComplex} will thus be proven.

So let $A,\vec{X},\vec{H}$ be an instance of the semilinear home-space
problem where $A$ is a Petri net, of dimension $d$, and
$\vec{X},\vec{H}$ are two (presentations of) semilinear subsets of
$\setN^d$. Since $\vec{H}$ can be decomposed, in elementary time, into a
finite union of linear sets using presentations with at most $d$
periods~\cite[Lemma~6.6]{GS64}, we can assume that each linear set
$\vec{L}$ of
the presentation of $\vec{H}$ satisfies this constraint. We put $n=d+\max\{\norm{A},\norm{\vec{X}},\norm{\vec{H}}\}$.

We first consider the problem of computing a semilinear
non-reachability core for each linear set $\vec{L}$ of the
presentation of $\vec{H}$. Such a linear set $\vec{L}$ is presented
with a basis $\vec{b}$ and a sequence of $k$ periods
$\vec{p}_1,\vec{p}_2,\ldots,\vec{p}_k$ with $k\leq d$. As previously
shown, this computation reduces to the computation of the minimal
elements of the upward closed set $\DCB$ and the upward-closed sets
$\vec{U}_{\vec{y}}$ where $\vec{y}$ belongs to the finite set of
vectors in $\setN^d$ satisfying $\norm{\vec{y}}\leq \norm{\vec{b}}$.
The computation of those minimal elements can be obtained by rewriting
the definitions of $\DCB$ and $\vec{U}_{\vec{y}}$ to match the
statement of Corollary~\ref{cor:pred}. To do so, we note that $\DCB$
and $\vec{U}_{\vec{y}}$ can be described in the following way:
\begin{align*}
  \DCB
  &=\{(\vec{y},\vec{u})\in\setN^d\times\setN^k \mid \exists \alpha,\beta\in\setN^d: \alpha\xrightarrow{A^*}\beta\wedge (\vec{y},\vec{u},\alpha,\beta)\in P\}
  \\
  \vec{U}_{\vec{y}}
  &=\{\vec{u}\in\setN^k \mid \exists \alpha,\beta\in\setN^d: \alpha\xrightarrow{A^*}\beta\wedge (\vec{u},\alpha,\beta)\in P_{\vec{y}}\}
\end{align*}
where:
\begin{align*}
  P&=\left\{(\vec{y},\vec{u},\alpha,\beta)\in\setN^d\times\setN^k\times\setN^d\times\setN^d\mid \exists (\vec{y}',\vec{u}')\in\setN^d\times\setN^k:\begin{array}{l}\norm{\vec{y}}>\norm{\vec{y}'}\wedge\\ \alpha=\prmark(\vec{y},\vec{u})\wedge\\ \beta=\prmark(\vec{y}',\vec{u}')\end{array}\right\}
  \\
  P_{\vec{y}}&=\{(\vec{u},\alpha,\beta)\in\setN^k\times\setN^d\times\setN^d
	\mid \alpha=\prmark(\vec{y},\vec{u}) \wedge \beta\in\vec{L}\}.
\end{align*}
Since the sets $P$ and $P_{\vec{y}}$ are clearly expressible by 
formulas in Presburger arithmetic, we can effectively construct, in
elementary time, semilinear presentations of those
sets~\cite{GS-PACIF66}. We introduce an elementary function $E$
(independent of any instance) corresponding to that computation. We
deduce that for some constant $c'>0$, independent of any input, we can
compute, in time exponential in $F_{8d+6}(c'E(n))$, the sets
$\min(\DCB)$ and $\min(\vec{U}_{\vec{y}})$ for
$\norm{\vec{y}}\leq\norm{\vec{b}}$. Moreover, the norms of vectors in
those sets are bounded by $F_{8d+6}(c'E(n))$. It follows from the
proof of Proposition~\ref{prop:coreforlinear} that there exists an
elementary function $E'$ (independent of any instance) such that we
can compute, in time $E'(F_{8d+6}(c'E(n)))$, a~(presentation of a)
semilinear non-reachability core $\vec{C}$ for each linear set $\vec{L}$ of the
presentation of $\vec{H}$.

Let $\vec{L}_1,\vec{L}_2,\ldots,\vec{L}_m$ be the presentation
sequence of $\vec{H}$, and let $\vec{C}_1,\vec{C}_2,\ldots,\vec{C}_m$
be the respective semilinear non-reachability cores
computed for $\vec{L}_1,\vec{L}_2,\ldots,\vec{L}_m$, respectively,
as shown in the previous paragraph.
Proposition~\ref{prop:coreexecution} shows that $\vec{H}$ is not a~home-space for $\vec{X}$ if, and only if, there is an execution
\begin{equation}\label{eq:badcomputrep}
  \vec{x}_0\reachA \vec{x}_1\reachA \vec{x}_2\cdots\reachA
  \vec{x}_m
\end{equation}
where $\vec{x}_0\in\vec{X}$, and $\vec{x}_i\in \vec{C}_i$ for each
$i\in[1,m]$. 

The existence of such an execution can be decided by
Proposition~\ref{prop:decsemilinseq}, by a reduction to the
reachability problem for a Petri net of a dimension that is elementary
in $\max\{d,m,n\}$.
Theorem~\ref{thm:reach} thus entails that 
the semilinear home-space problem is decidable
in Ackermannian time, which finishes the proof of
Theorem~\ref{th:AckerComplex}.

\section{Semilinear Inductive Cores for Semilinear Sets}\label{sec:semilcore}
In Lemma~\ref{lem:coreforlinear} we proved that for any Petri net $A$
of dimension $d$ and (a presentation
of) a~linear set $\vec{L}\subseteq \setN^d$ there is an effectively constructible 
semilinear non-reachability core $\vec{C}$ for $\vec{L}$. 
A natural question is if we can compute a semilinear core for any
semilinear set.
By Proposition~\ref{prop:union}, this is the case if we can extend
Lemma~\ref{lem:coreforlinear} so that the respective semilinear cores $\vec{C}$ for
linear sets $\vec{L}$ are, moreover, inductive. We can indeed achieve
this,  by using the following known result and its corollary.
\begin{thmC}[{\cite[Theorem 8.3]{DBLP:journals/corr/abs-1009-1076}}]\label{thm:leroux} %FIXED Removed double Brackets
  Given a Petri net $A$ and two semilinear sets $\vec{X},\vec{Y}$ of configurations, we have 
  $\vec{X}\subseteq \overline{\pre(\vec{Y})}$ if, and only if, there exists an effectively 
  constructible semilinear inductive set $\vec{I}$ such that
  $\vec{X}\subseteq\vec{I}\subseteq\overline{\vec{Y}}$. 
\end{thmC}
\begin{cor}\label{cor:leroux}
 Given a Petri net $A$ and two semilinear sets $\vec{C},\vec{H}$ of
	configurations where $\vec{C}$ is a non-reachability core for
	$\vec{H}$, there is an effectively 
  constructible semilinear inductive
	non-reachability core $\vec{C}'\supseteq\vec{C}$ for
	$\vec{H}$.
\end{cor}	
\begin{proof}
	For the considered $A,\vec{C},\vec{H}$ we have
	$\vec{C}\subseteq\overline{\pre(\vec{H})}\subseteq\pre(\vec{C})$.
	By Theorem~\ref{thm:leroux} there is 
an effectively 
  constructible semilinear inductive set
	$\vec{C}'$ such that 
	$\vec{C}\subseteq\vec{C}'\subseteq\overline{\vec{H}}$.
	Since $\vec{C}'$ is inductive, i.e.\
	$\post(\vec{C}')=\vec{C'}$, we also have 
	$\pre(\overline{\vec{C}'})=\overline{\vec{C}'}$.
Hence $\vec{C}'\subseteq\overline{\vec{H}}$, i.e.\  
$\vec{H}\subseteq\overline{\vec{C}'}$, 
	entails
	$\pre(\vec{H})\subseteq\pre(\overline{\vec{C}'})=\overline{\vec{C}'}$, 
	i.e.\ $\vec{C}'\subseteq\overline{\pre(\vec{H})}$.
	We thus have
	$\vec{C}\subseteq\vec{C}'\subseteq\overline{\pre(\vec{H})}\subseteq\pre(\vec{C})\subseteq\pre(\vec{C}')$.
\end{proof}
%NEEDFIX
We can thus deduce the following theorem.
\begin{thm}\label{thm:core}
  Given a Petri net $A$ of dimension $d$, and (a presentation
  of) a~semilinear set
  $\vec{H}\subseteq \setN^d$, there is an effectively constructible 
  semilinear inductive non-reachability core $\vec{C}$ for $\vec{H}$.
\end{thm}
\begin{proof}
  If $\vec{H}=\emptyset$, then we can take $\vec{C}=\setN^d$.
  
  Now we assume that $\vec{H}=\vec{H}_1\cup
  \vec{H}_2\cdots\cup \vec{H}_m$ for some $m\geq 1$, where
  $\vec{H}_i$ is a~linear set for each $i\in[1,m]$.
  By Lemma~\ref{lem:coreforlinear} we can construct semilinear sets
  $\vec{C}_1,\vec{C}_2,\dots,\vec{C}_m$ that are
  non-reachability cores for
  $\vec{H}_1,\vec{H}_2,\dots,\vec{H}_m$, respectively
  (hence $\vec{C}_i\subseteq 
  \overline{\pre(\vec{H}_i)}\subseteq\pre(\vec{C}_i)$).
  
	By Corollary~\ref{cor:leroux},
for each $i\in[1,m]$ we can
	construct a semilinear inductive core $\vec{C}'_i$ for
	$\vec{H}_i$. 
    By Proposition~\ref{prop:union} we deduce that 
  $\vec{C}'_1\cap\vec{C}'_2\cdots\cap\vec{C}'_m$ is a semilinear
  inductive non-reachability core for $\vec{H}$
  (using the fact that the intersection of semilinear sets
  is effectively semilinear).
\end{proof}

\begin{rem}
	Theorem~\ref{thm:core} also yields the decidability of the semilinear
	home-space problem, since the problem if
	$\vec{X}\reachA\vec{C}$ (i.e., if $\vec{x}\reachA\vec{c}$ for
	some $\vec{x}\in\vec{X}$ and $\vec{c}\in\vec{C}$)
	for semilinear sets $\vec{X},\vec{C}$ of configurations of a
	given
	Petri net is decidable; we have already recalled that 
	the semilinear reachability
	problem is easily reducible to the standard
	reachability problem, thus being also Ackermann-complete.
	Theorem~\ref{thm:leroux}
	from~\cite{DBLP:journals/corr/abs-1009-1076} gives us no
	complexity bound for constructing the inductive semilinear set
	$\vec{I}$; therefore we could not derive any complexity bound in
	this way. In fact, this would be possible now; we could 
	show that the inductive semilinear cores for semilinear sets
	are computable in Ackermannian time, 
        by using the results of a new paper~\cite{DBLP:conf/fossacs/Leroux24}.
Such a complexity proof would be thus based 
	on an involved result about semilinear inductive invariants,
        whereas the complexity proof presented in this paper 
is independent of this.
\end{rem}	

\section{Home-Space Witnesses}\label{sec:witnesses}

We recall that the reachability problem for Petri nets is decidable
but extremely hard, namely Ackermann-complete. 
Nevertheless there are positive witnesses of
reachability that are easily verifiable: given a Petri net $A$ and two
configurations $\vec{x},\vec{y}$, a witness of the fact
$\vec{x}\reachA\vec{y}$ is simply a word $w\in A^*$ such that
$\vec{x}\xrightarrow{w}\vec{y}$.
Verifying the validity of $\vec{x}\xrightarrow{w}\vec{y}$ is trivial;
of course, the size of such a witness $w$ is another issue.
A negative witness, meaning a witness of the fact
$\vec{x}\not\reachA\vec{y}$, is a~more involved question; a solution is
provided by Theorem~\ref{thm:leroux}:  we have $\vec{x}\not\reachA\vec{y}$
if{f} there is an inductive semilinear $\vec{I}$ such that
$\vec{x}\in\vec{I}$ and $\vec{y}\not\in\vec{I}$.
Verifying if a~given semilinear set $\vec{I}$ is inductive and
satisfies $\vec{x}\in\vec{I}$ and $\vec{y}\not\in\vec{I}$
 is much easier than solving
the reachability problem
(we can refer, e.g., to~\cite{DBLP:journals/siglog/Haase18} for
complexity details);
again, the size of such a witness $\vec{I}$ is another issue.

When looking for similar witnesses in the case of the semilinear home-space
problem, the following lemma provides a solution in terms of semilinear inductive invariants.
\begin{lem}\label{lem:middleS}
Given a~Petri net $A$ of dimension $d$, 
  and two semilinear sets $\vec{X},\vec{H}\subseteq\setN^d$,
 we have $\post(\vec{X})\subseteq\pre(\vec{H})$
  (i.e., $\vec{H}$ is a
  home-space for $\vec{X}$) if{f}  there
  is an inductive semilinear set $\vec{I}$ such that
$\post(\vec{X})\subseteq \vec{I}\subseteq\pre(\vec{H})$.
\end{lem}
\begin{proof}
  The ``if'' direction is trivial.
  
  Now we show the ``only if'' direction. Let us  
  assume that 
  $\post(\vec{X})\subseteq\pre(\vec{H})$, and let $\vec{C}$
  be a semilinear non-reachability core for $\vec{H}$ guaranteed by
	Theorem~\ref{thm:core} 
	(while here we do not need $\vec{C}$ to be inductive); we thus have 
	$\vec{C}\subseteq\overline{\pre(\vec{H})}\subseteq\pre(\vec{C})$.
The assumption 
  $\post(\vec{X})\subseteq\pre(\vec{H})$ thus entails that 
  $\vec{X}\cap\pre(\vec{C})=\emptyset$, i.e.,
  $\vec{X}\subseteq\overline{\pre(\vec{C})}$.
  Hence by Theorem~\ref{thm:leroux} there is an inductive
  semilinear set $\vec{I}$ such that
  $\vec{X}\subseteq\vec{I}\subseteq \overline{\vec{C}}$.
	Since $\vec{I}$ is inductive (hence $\post(\vec{I})=\vec{I}$
	and $\pre(\overline{\vec{I}})=\overline{\vec{I}}$), 
$\vec{X}\subseteq\vec{I}$ entails
$\post(\vec{X})\subseteq\vec{I}$,
and $\vec{I}\subseteq \overline{\vec{C}}$, i.e.\ 
	$\vec{C}\subseteq \overline{\vec{I}}$, entails
	$\pre(\vec{C})\subseteq \overline{\vec{I}}$, i.e.\ 
		$\vec{I}\subseteq \overline{\pre(\vec{C})}$;
		moreover,
		$\overline{\pre(\vec{H})}\subseteq\pre(\vec{C})$
		entails
		$\overline{\pre(\vec{C})}\subseteq\pre(\vec{H})$,
		hence 	$\vec{I}\subseteq \pre(\vec{H})$.
\end{proof}
\noindent %FIXED indent
Let us look at the question of verifying the validity of a witness
$\vec{I}$ suggested by Lemma~\ref{lem:middleS}.
Given a Petri net $A$ and two semilinear sets 
 $\vec{X},\vec{H}$ of its configurations, for a given semilinear
 $\vec{I}$ we can ``easily'' (see~\cite{DBLP:journals/siglog/Haase18})
decide if $\vec{I}$
 is inductive and subsumes $\vec{X}$ (which entails that
 $\post(\vec{X})\subseteq \vec{I}$). For deciding if
 $\vec{I}\subseteq\pre(\vec{H})$ we also try to avoid solving the (semilinear)
 reachability problem; we achieve this by the following extension of
 (positive) witnesses
 $\vec{I}$.

Given a $d$-dimensional Petri net $A$, a \emph{positive home-space
witness for a pair} $(\vec{X},\vec{H})$ of semilinear subsets of $\setN^d$
is a pair 
\[(\vec{I},(w_1,w_2,\dots,w_k))\]
(for some $k\in\setN$)
where $\vec{I}\subseteq\setN^d$ is an inductive semilinear set that
contains $\vec{X}$, and $w_1,w_2,\ldots,w_k$ are words
from $A^+$ 
satisfying the following formula:
\begin{equation}\label{eq:positivewitness}
	(\forall \vec{y}\in \vec{I})(\exists
	n_1,n_2,\ldots,n_k\in\setN)(\exists \vec{h}\in \vec{H})
	\,\vec{y}\xrightarrow{w_1^{n_1}w_2^{n_2}\cdots
	w_k^{n_k}}\vec{h}.
\end{equation}
From~\cite[Theorem XIII.2]{DBLP:conf/lics/Leroux13} we deduce 
that there exists a sequence $(w_1,w_2,\ldots,w_k)$ satisfying~(\ref{eq:positivewitness})
precisely when $\vec{I}\subseteq\pre(\vec{H})$ (since $\vec{I}$ and
$\vec{H}$ are semilinear).

\begin{cor}
  Given a~Petri net $A$ 
  and two semilinear sets $\vec{X},\vec{H}$ of its configurations,
  the set  $\vec{H}$ is a
  home-space for $\vec{X}$ if{f}  there is a positive home-space
  witness $(\vec{I},(w_1,w_2,\dots,w_k))$ for
  $(\vec{X},\vec{H})$.
\end{cor}
\noindent %FIXED indent
We note that by compiling the relation
$\vec{y}\xrightarrow{w_1^{n_1}w_2^{n_2}\cdots w_k^{n_k}}\vec{h}$ into
a Presburger formula over the free variables
$\vec{y},n_1,\ldots,n_k,\vec{h}$~(see
\cite{DBLP:conf/concur/FribourgO97} for details), we deduce that
formula~(\ref{eq:positivewitness}) can be
efficiently
transformed into a
Presburger formula. Since the complexity of Presburger arithmetic is
at most 3-exponential~\cite{DBLP:journals/jcss/Oppen78},
checking if a tuple $(\vec{I},(w_1,w_2,\dots,w_k))$ is a
positive home-space witness for a~pair $(\vec{X},\vec{H})$
is elementary (while 
the general reachability problem is nonelementary, namely
Ackermann-complete).

\begin{rem}
  A negative home-space witness for a pair $(\vec{X},\vec{H})$
  of semilinear sets of configurations, which exists precisely
  when $\vec{H}$ is not a home space for $\vec{X}$, can be
  defined as a tuple $(\vec{x},\vec{y})$ where
  $\vec{x}\in\vec{X}$, $\vec{x}\reachA\vec{y}$, and
  $\vec{y}\not\reachA\vec{H}$. To avoid requirements to solve
  instances of
  the reachability problem, we can define such a~negative
  witness as a tuple $(\vec{Y},\vec{x},w,\vec{y})$ where 
  $\vec{Y}$ is an inductive semilinear set disjoint from
  $\vec{H}$, and we have $\vec{x}\in \vec{X}$, $\vec{y}\in \vec{Y}$, 
  and	$\vec{x}\xrightarrow{w}\vec{y}$.
  
  Hence deciding if $\vec{H}$ is a home-space for $\vec{X}$ can
  be performed by simultaneously searching for a positive or a
  negative witness. This also yields the decidability of the
  semilinear home-space problem.
\end{rem}

\section{Concluding Remarks}\label{sec:conclusions}
There are various issues that can be elaborated on and added to
the presented material.
One such issue was mentioned in Remark~\ref{rem:fixedfivedim}, dealing with 
strengthening the lower bound. 

We also leave open the complexity of deciding if an inductive
semilinear set $\vec{C}$ is a~non-reachability core for a semilinear set $\vec{H}$. Let us recall that this problem is equivalent to prove that $\vec{C}$ is disjoint from $\vec{H}$, and $\pre(\vec{C}\cup\vec{H})$ is the full set of configurations $\setN^d$. This last problem is related to the \emph{semilinear universal problem} for Petri nets defined as follows:
\begin{quote}
	\emph{Instance:} a Petri net $A$, of dimension $d$, 
	and a semilinear set $\vec{S}\subseteq\setN^d$.
\\
	\emph{Question:} is
	$\post(\vec{S})=\setN^d$~?
\end{quote}
The paper~\cite{DBLP:journals/fuin/JancarLS19} shows that the problem is decidable, and expspace-complete when the problem is restricted to singleton sets $\vec{S}$. In general, the complexity of the problem is still open.

Best and Esparza~\cite{DBLP:journals/ipl/BestE16}
consider the ``existential'' home-space problem that asks, given a~Petri net $A$ of dimension $d$
and an initial configuration $\vec{x}$,
if there exists a singleton home-space 
for $\{\vec{x}\}$; the main result of~\cite{DBLP:journals/ipl/BestE16}
shows that this existential problem is decidable.
We can consider a related problem that asks, given $A$
and $\vec{x}$, if there is 
a semilinear home-space included in $\post(\vec{x})$; currently we have no
answer to the
respective decidability question.

\bibliographystyle{alphaurl}
\bibliography{bibliography}

\end{document}